\documentstyle[12pt,epsf]{article}
\newcommand{\beq}{\begin{equation}}
\newcommand{\eeq}{\end{equation}}
\newcommand{\beqa}{\begin{eqnarray}}
\newcommand{\eeqa}{\end{eqnarray}}

\newcommand{\krig}[1]{\stackrel{\circ}{#1}}

\newcommand{\no}{\nonumber}

\begin{document}

\hfill TK 96 14

\hfill hep-ph/9607432

\bigskip\bigskip

\begin{center}

{{\Large\bf Chiral expansion of baryon masses and $\sigma$--terms}}

\end{center}

\vspace{.4in}

\begin{center}
{\large B. Borasoy\footnote{email: borasoy@itkp.uni-bonn.de},
 Ulf-G. Mei{\ss}ner\footnote{email: meissner@itkp.uni-bonn.de}\footnote{
Address after Oct. 1, 1996: FZ J\"ulich, IKP (Theorie), D-52425 J\"ulich,
Germany}}

\bigskip

\bigskip

Universit\"at Bonn, Institut f{\"u}r Theoretische Kernphysik\\
Nussallee 14-16, D-53115 Bonn, Germany\\
\end{center}

\vspace{.7in}

\thispagestyle{empty} 

\begin{abstract}
We analyze the octet baryon masses and the pion/kaon--nucleon
$\sigma$--terms in the framework of heavy baryon chiral 
perturbation theory. We include {\it all} terms up-to-and-including 
quadratic order in the light quark masses, $m_q$. 
We develop a consistent scheme to estimate low--energy constants
related to scalar--isoscalar operators in the framework of resonance
exchange involving  one--loop graphs. The pertinent low--energy 
constants  can only be estimated up to some finite coefficients.
Including contributions from 
loop graphs with intermediate spin--3/2 decuplet and
spin--1/2 octet states and from tree graphs including scalar
mesons, we use the octet baryon masses and the pion--nucleon $\sigma$--term
to fix all but one of these coefficients. Physical results are
insensitive to this remaining parameter. 
It is  also demonstrated that  two--loop corrections only modify
some of the subleading low--energy constants. 
We find for the baryon mass in the chiral limit,  $\krig{m} =
770 \pm 110$~MeV. While the  corrections of order $m_q^2$ are small
for the nucleon, they are still sizeable for the $\Lambda$, the
$\Sigma$ and the $\Xi$. Therefore a definitive statement about the
convergence of three--flavor baryon chiral perturbation can not yet be
made. The  strangeness content of the nucleon is $y = 0.21 \pm
0.20$. We also estimate the kaon--nucleon $\sigma$--terms and some 
two--loops contributions to the nucleon mass. 

\end{abstract}

\vfill


\section{Introduction}
Despite two decades of calculations, there is still sizeable
uncertainty about the quark mass expansion
of the octet baryon masses in the framework of chiral
perturbation theory, see e.g. 
\cite{lang,pagels,juerg,gl82,liz,dobo,bkmz,lelu,samir,mary,dent,buergi,ak}. 
The penultimate goal of these calculations is to get additional bounds
on the ratios of the light quark masses and to understand the success
of the Gell--Mann--Okubo mass relation. Originally, it was argued
that the leading non--analytic corrections of the type $M_\phi^3$,
where $\phi$ denotes  the Goldstone boson fields $\pi, K, \eta$, are
so large that chiral perturbation theory is meaningless in this case
(for the most extensive study see Gasser \cite{juerg}). In the same
paper, a meson--cloud model was proposed which showed a faster
convergence and allowed to give some bounds on the quark mass ratios.
With the advent of a heavy baryon chiral perturbation theory  as
proposed by Jenkins and Manohar \cite{jm}, which allows for a
consistent power counting and is in spirit similar to the meson--cloud
model, the problem of the chiral expansion of the baryon masses was taken
up again. In particular, it was argued by Jenkins \cite{liz} that the
inclusion of the decuplet in the effective field theory provides a
natural framework to account for the large cancellations
between meson loop and contact term contributions at orders $q^2$, $q^3$
and (partly) $q^4$ (where $q$ denotes a small momentum or meson mass)
thus giving credit
to the Gell--Mann--Okubo (GMO) mass relation.  However, the inclusion of the
decuplet in that approach was not done systematically and criticized
in Ref.\cite{bkmz}.\footnote{Notice that only recently a systematic
effective field theory formulation for spin--3/2 fields has become 
available for the two--flavor case \cite{hhk}.} Lebed and Luty
\cite{lelu} analyzed the baryon masses to second order in the quark
masses, claiming that the quark mass ratios and deviations from the  
GMO relations are not computable, whereas the corrections to the 
Coleman--Glashow relations in the decuplet are. We have already
pointed out in the letter \cite{bmlet} that in that paper not all
contributions at order $m_q^2$ and $m_q^2 \, \ln \, m_q$
are accounted for. Furthermore, most of the
existing papers considered mostly the so--called computable
corrections of order $m_q^2$ (modulo logs) or 
included some of the finite terms at this order \cite{liz,dobo,lelu,mary}.
This, however, contradicts the spirit of chiral perturbation 
theory (CHPT) in that all terms at a given order have to be retained, see
e.g. \cite{wein79,gl84,gl85}. 
In Ref.\cite{bmlet} we presented
the results of a first calculation including {\it all} terms of order
${\cal O}(m_q^2)$, where $m_q$ is a generic symbol for any one of the light
quark masses $m_{u,d,s}$. In that paper, the isospin limit 
$m_u = m_d$ was taken  and electromagnetic corrections were neglected.
Concerning the latter, we remark here that there is a fundamental
problem in that the electromagnetic interactions lead to an infinite 
renormalization of the quark masses as first pointed out by Gasser and
Leutwyler \cite{gl75}. This problem has yet to be addressed in a
systematic fashion and we thus will work in a world with $\alpha=0$,
with $\alpha = e^2/4 \pi$ the fine structure constant. 
We note that the treatment of estimating low--energy constants from
intermediate decuplet states in Ref.\cite{bmlet} was inconsistent.
Here, we will give the correct and much more detailed treatment of 
the s--channel resonance exchanges to the
low--energy constants and thus refine our previous analysis \cite{bmlet}.

Closely related to the chiral expansion of the baryon masses is the 
pion--nucleon $\sigma$--term \cite{juerg,bkmz,jms} (and also the
two kaon--nucleon $\sigma$--terms, $\sigma_{KN}^{(1,2)}$). These
$\sigma$--terms measure the strengths of the scalar quark condensates
in the proton and vanish in the chiral limit of zero quark
masses. They are thus particularly suited to test our understanding of
the mechanism of the spontaneous and explicit chiral symmetry
breaking. Again, at present
no clear picture concerning the quark mass expansion of these
quantities has emerged, although it is believed that the
dispersion--theoretical analysis of Gasser, Leutwyler and Sainio
\cite{gls} has significantly sharpened our understanding of relating
$\sigma_{\pi N}$ as extracted from the
pion--nucleon scattering data to the expectation value
of $\hat{m}(\bar u u + \bar d d)$ in the proton, where $\hat m = (m_u
+m_d)/2$ denotes the average light quark mass. Of particular interest
is the so--called strangeness content of the proton, i.e. the matrix
element $<p|m_s \bar{s} s |p>$. It can be indirectly inferred from the
analysis of the pion--nucleon $\sigma$--term. The present status can
be summarized by the statement that this matrix element is
non--vanishing but also not particularly large, as one deduces
from the lowest order quark mass analysis \cite{juerg} (for a status 
report, see e.g. Ref.\cite{mikko}).
Our aim is to show to what accuarcy one can at present make statements
about the quark mass expansion of the $\sigma$--terms. For some 
comprehensive reviews, see e.g.  \cite{ulfrev,ecker,bkmrev,toni}.

The paper is organized as follows. In section \ref{sec:EFFLA} we write
down the effective meson--baryon Lagrangian, i.e. all terms which
contribute to the baryon masses and the $\sigma$--terms. The minimal
number of contact terms up--to--and--including ${\cal O}(q^4)$
consists of seven terms of dimension two and seven terms of dimension
four. These terms are accompanied by so--called low--energy
constants (LECs), i.e. coupling constants not fixed by chiral symmetry.
We remark that some of the dimension  four LECs can be absorbed in
some of the dimension two ones since they amount to quark mass
renormalizations of the latter. This ambiguity is also discussed.
In section \ref{sec:massigren} we give the formulae for the fourth
order contribution to the octet baryon masses and the pion--nucleon
$\sigma$--term. We also discuss the renormalization to render the
divergent loop diagrams at order $q^4$ finite. This is in marked
contrast to the calculation of the leading non--analytic corrections
which are all finite. Section \ref{sec:LECS} deals with the estimate
of the fourteen LECs from resonance exchange. We consider contributions
from the spin--3/2 decuplet, the spin--1/2 octet of even-parity
excited baryon resonances and from t--channel scalar meson exchange.
Considering loop graphs with intermediate resonances leads to
divergences to leading order. The corresponding renormalization
constants can not be fixed from resonance parameters. This problem
was incorrectly treated in \cite{bmlet}. However, using the baryon
masses and the pion--nucleon $\sigma$--term, we are able to fix all
but one of these coefficients. We show that the corresponding LECs and
physical observables are essentially insensitive to the choice of this
parameter which is bounded phenomenologically.
We also address the question of two (and higher) loop contributions
to the LECs. This is mandated by the fact that the estimation of the
LECs for the baryon masses and $\sigma$--terms involves Goldstone
boson loops (treated in a particular fashion as explained in 
section~\ref{sec:LECS}) and intermediate resonance propagators. Therefore, a
non--trivial reodering of the chiral expansion for such graphs is
expected. As we will show  two--loop graphs only contribute to
some of the subleading LECs at the same order as the one--loop
diagrams. Most of the two--loop contributions can be completely
absorbed in a redefinition of the one loop renormalization parameters.
The results  are given in section \ref{sec:results} including a
detailed study of the dependence on certain input parameters. A short
summary and outlook is given in section \ref{sec:summary}. Some
lengthy formulae and technicalities are relegated to the appendices.


\section{Effective Lagrangian}
\label{sec:EFFLA}

To perform the calculations, we make use of the effective
meson--baryon Lagrangian. To construct the various terms, we start
from the relativistic formulation which is then reduced to the heavy
fermion limit. This has the advantage of automatically generating all 
kinematical $1/m$ corrections with the correct coefficients, see e.g.
appendix~A of Ref.\cite{bkkm}. However, we do not spell out the
relativistic Lagrangian but directly the heavy mass form emerging from
it. In the extreme non--relativistic limit \cite{jm,bkkm}, the baryons
are characterized by a four--velocity $v_\mu$. In this approach, there
is a one--to--one
correspondence between the expansion in small momenta and quark masses
and the expansion in Goldstone boson loops, i.e. a consistent power
counting scheme emerges. Our notation is identical to the one
used in \cite{bkmz} and we discuss here only the new terms necessary for
the calculations of the masses and $\sigma$--terms.
The pseudoscalar Goldstone fields ($\phi = \pi, K, \eta$) are collected in
the  $3 \times 3$ unimodular, unitary matrix $U(x)$, 
\begin{equation}
 U(\phi) = u^2 (\phi) = \exp \lbrace i \phi / F_\pi \rbrace
\end{equation}
with $F_\pi$ the pseudoscalar decay constant (in the chiral limit), and
\begin{eqnarray}
 \phi =  \sqrt{2}  \left(
\matrix { {1\over \sqrt 2} \pi^0 + {1 \over \sqrt 6} \eta
&\pi^+ &K^+ \nonumber \\
\pi^-
        & -{1\over \sqrt 2} \pi^0 + {1 \over \sqrt 6} \eta & K^0
        \nonumber \\
K^-
        &  \bar{K^0}&- {2 \over \sqrt 6} \eta  \nonumber \\} 
\!\!\!\!\!\!\!\!\!\!\!\!\!\!\! \right) \, \, \, \, \, . 
\end{eqnarray}
Under SU(3)$_L \times$SU(3)$_R$, $U(x)$ transforms as $U \to U' =
LUR^\dagger$, with $L,R \in$ SU(3)$_{L,R}$.
The matrix $B$ denotes the baryon octet, 
\begin{eqnarray}
B  =  \left(
\matrix  { {1\over \sqrt 2} \Sigma^0 + {1 \over \sqrt 6} \Lambda
&\Sigma^+ &  p \nonumber \\
\Sigma^-
    & -{1\over \sqrt 2} \Sigma^0 + {1 \over \sqrt 6} \Lambda & n
    \nonumber \\
\Xi^-
        &       \Xi^0 &- {2 \over \sqrt 6} \Lambda \nonumber \\} 
\!\!\!\!\!\!\!\!\!\!\!\!\!\!\!\!\! \right)  \, \, \, .
\end{eqnarray}
Under $SU(3)_L \times SU(3)_R$, $B$  transforms as any matter field,
\begin{equation} 
B \to B' = K \, B \,  K^\dagger
 \, \, \, ,
\end{equation}
with $K(U,L,R)$ the compensator field representing an element of the
conserved subgroup SU(3)$_V$. The effective Lagrangian takes the form
\begin{equation}
{\cal L}_{\rm eff} = {\cal L}_{\phi B}^{(1)} +  {\cal L}_{\phi B}^{(2)} +
  {\cal L}_{\phi B}^{(3)} +
 {\cal L}_{\phi B}^{(4)} + {\cal L}_{\phi}^{(2)}+ {\cal L}_{\phi}^{(4)}
\label{leff}
\end{equation}
where the chiral dimension $(i)$ counts the number of derivatives
and/or meson mass insertions. Before discussing the various terms in 
${\cal L}_{\rm eff}$, a few words about the power counting are in
order. Clearly, derivatives (momenta) and meson masses count as order
$q$. For the quark masses, we stick to the standard scenario, which
prescribes a power $q^2$ to scalar and pseudoscalar external sources,
in particular to the eigenvalues of the quark mass matrix ${\cal M}$.
It would certainly be interesting to study the alternative approach in
which the quark masses are counted as order $q$ only. Clearly, one
then would have to deal with much more terms at a given fixed order.
We do not attempt such a 'generalized' analysis here. We remark that
the dimension three operators in  ${\cal L}_{\phi B}^{(3)}$ do not
contribute directly to the masses but are needed for the
renormalization of the baryon self--energy.

The form of the lowest order meson--baryon Lagrangian is
\begin{equation} 
{\cal L}_{\phi B}^{(1)} = i \, \bar{B} \,  v \cdot D \,  B 
+ F \, {\rm Tr}(\bar{B} S_\mu \, [u^\mu , B])+ D \, {\rm Tr}(\bar{B} 
S_\mu \, \{u^\mu ,B\}) \, \, \, \, ,
\label{LMB1}
\end{equation}
where $D \simeq 3/4$ and $F \simeq 1/2$ are the two axial--vector
coupling constants, $D_\mu = \partial_\mu + \Gamma_\mu$ is the chiral
covariant derivative\footnote{For the calculations performed here, we
only need the partial derivative in $D_\mu$.} 
and $S_\mu$ denotes the covariant spin--operator
{\`a} la Pauli--Lubanski.  The meson Lagrangian 
${\cal L}_{\phi}^{(2)}+ {\cal L}_{\phi}^{(4)}$ is given in \cite{gl85}. 

The dimension two meson--baryon Lagrangian can be written as
(we only enumerate the terms which contribute)
\begin{equation}
{\cal L}_{\phi B}^{(2)} = {\cal L}_{\phi B}^{(2, {\rm br})} +
\sum_{i=1}^{4} \, b_i \, O_i^{(2)} + 
{\cal L}_{\phi B}^{(2, {\rm rc})}\, \, ,   
\label{leff2}
\end{equation}
with the $O_i^{(2)}$ monomials in the fields of chiral dimension two
discussed below. The explicit symmetry breaking terms are collected 
in ${\cal L}_{\phi B}^{(2, {\rm br})}$  \cite{bkmz},
\begin{equation}
{\cal L}_{\phi B}^{(2, {\rm br})} = b_D \, {\rm Tr}(\bar B \lbrace
\chi_+ , B \rbrace ) + b_F \, {\rm Tr}(\bar B [\chi_+,B]) + 
b_0 \, {\rm Tr}(\bar B
B) \, {\rm Tr}(\chi_+ ) \, \, ,   
\label{leff2st}
\end{equation}
i.e. it contains three low--energy constants $b_{0,D,F}$.
$\chi_+ = u^\dagger
\chi u^\dagger + u \chi^\dagger u$ is proportional to the quark mass
matrix ${\cal M} ={\rm diag}(m_u,m_d,m_s)$ since $\chi = 2 B {\cal
  M}$. Here, $B = - \langle 0 | \bar{q} q | 0 \rangle / F_\pi^2$ is
the order parameter of the spontaneous symmetry violation. 
We assume $B \gg F_\pi$. In what
follows, we will work to order $q^4$ and thus have to include terms
with derivatives on the Goldstone boson fields. The minimal set of such terms
is given by
\begin{eqnarray}
\sum_{i=1,2,3,8} \, b_i \, O_i^{(2)} & = & 
b_1 \, {\rm Tr}(\bar{B} [ u_\mu , [ u^\mu,B   ]]) +
b_2 \, {\rm Tr}(\bar{B} [ u_\mu , \{ u^\mu,B \}]) \nonumber \\
& + &
b_3 \, {\rm Tr}(\bar{B} \{ u_\mu , \{ u^\mu,B \} \}) +
b_8 \, {\rm Tr}(\bar{B}B ) {\rm Tr}(u^\mu u_\mu) 
\label{leff2der}
\end{eqnarray}
with $u_\mu = i  u^\dagger \partial_\mu U u^\dagger$.
A couple of remarks are in order. First, there are also terms of the type
$\bar{B}(v \cdot u)^2 B$. Because of the identity (in $d$ 
space--time dimensions)
\begin{equation}
\int d^dk \frac{(v \cdot k)^2}{k^2 - M_\pi^2} =
\frac{1}{d} \,\int d^dk \frac{ k^2}{k^2 - M_\pi^2} \, \, \, \, ,
\end{equation} 
which holds for all the graphs to be considered later on, the
respective coupling constants can completely be absorbed in the
$b_{1,2,3,8}$. Second, we heavily  make use of the Cayley--Hamilton
identity for any traceless $3 \times 3$ matrix $X$,
\beq
{\rm Tr}(\bar{B} \{ X^2 , B\}) + {\rm Tr}(\bar{B}XBX) -\frac{1}{2}
{\rm Tr}(\bar{B}B){\rm Tr}(X^2) - {\rm Tr}(\bar{B}X){\rm Tr}(BX) = 0
\, \, \, \, \, .
\label{CaHa}
\eeq
This allows e.g. to eliminate the term of the type
Tr$(\bar{B}u_\mu)$Tr$(u^\mu B)$ in terms of the $O_1$, $O_3$ and $O_8$,
i.e.
\begin{eqnarray}
\frac{1}{4}{\rm Tr}(\bar{B} [u_\mu,[u^\mu,B]]) &+& 
\frac{3}{4}{\rm Tr}(\bar{B} \{u_\mu, \{ u^\mu ,B \} \} )
\nonumber \\ & &-\frac{1}{2}
{\rm Tr}(\bar{B}B){\rm Tr}(u_\mu u^\mu) - {\rm Tr}(\bar{B}u_\mu )
{\rm Tr}(u^\mu B) = 0
\, \, \, \, \, .
\end{eqnarray}
The basis used in Ref.\cite{bmlet} was overcomplete and thus a larger
number of terms was counted. This does, however, not directly matter
for the calculations.\footnote{Notice that for facilitating the
comparison with the results of Ref.\cite{bmlet}, we keep the numbering of
the terms used there.}
There are also relativistic corrections of dimension two, i.e. $1/m$
corrections with fixed coefficients. These read
\begin{eqnarray}
{\cal L}_{\phi B}^{(2, {\rm rc})} & = & 
-\frac{iD}{ 2 \krig{m}} {\rm Tr}(\bar{B}S_\mu [D^\mu, \{v \cdot u,B
\}])
-\frac{iF}{ 2 \krig{m}} {\rm Tr}(\bar{B}S_\mu [D^\mu, [v \cdot
u,B]]) \nonumber \\
& & 
-\frac{iF}{ 2 \krig{m}} {\rm Tr}(\bar{B}S_\mu [v \cdot u , [
D^\mu, B ]\})
-\frac{iD }{ 2 \krig{m}} {\rm Tr}(\bar{B}S_\mu \{ v \cdot u, [D^\mu, 
B] \} ) \nonumber \\ 
& & 
+\frac{D^2-3F^2}{24\krig{m}} {\rm Tr}(\bar{B} [v \cdot u, [v \cdot u , B ]])
-\frac{D^2}{12 \krig{m}} {\rm Tr}(\bar{B} B ) \, {\rm Tr}(v \cdot u \,
v \cdot u ) \nonumber \\ & & 
-\frac{1}{2\krig m}\, {\rm Tr}(\bar B \,[D_\mu, [D^\mu,B]])
+\frac{1}{2\krig m}\, {\rm Tr}(\bar B \,[v \cdot D, [v \cdot D,B]])
\nonumber \\ & &
-\frac{DF}{ 4 \krig{m}} {\rm Tr}(\bar{B} [v \cdot u, \{ v \cdot u,B \}])  
\,\,\, ,  
\label{relcorr}
\end{eqnarray}
where $\krig{m}$ denotes the average baryon octet mass in the chiral limit.
All low--energy constants in ${\cal L}_{\phi B}^{(2)}$ are finite.
The splitting of the dimension two meson--baryon Lagrangian in 
Eq.(\ref{leff2}) is motivated by the fact that while the first three
terms appear in tree and loop graphs, the latter eleven only come in via
loops. We remark that the LECs $b_i$ ($i=0,1,2,3,8,D,F$) have
dimension mass$^{-1}$. Notice also that the last three terms in 
Eq.(\ref{relcorr}) could be absorbed in the LECs $b_{1,2,3}$. For our
later resonance saturation estimates, we prefer to keep them separately.

There are seven independent terms contributing at dimension four,
which can be deduced from the over--complete basis
\begin{eqnarray}
{\cal L}_{\phi B}^{(4)} & = & 
\sum_{i=1}^{8} \, d_i \, O_i^{(4)}  
\nonumber \\
& = & d_1 \, {\rm Tr} (\bar{B} [\chi_+ , [ \chi_+ , B]] )   
 + d_2 \, {\rm Tr} (\bar{B} [\chi_+ , \{ \chi_+ , B\} ] ) \nonumber \\   
& + & d_3 \, {\rm Tr} (\bar{B} \{ \chi_+ , \{ \chi_+ , B\} \} )   
 + d_4 \, {\rm Tr}(\bar{B} \chi_+ ) {\rm Tr}(\chi_+ B) \nonumber \\
& + & d_5 \, {\rm Tr} (\bar{B}  [ \chi_+ , B] ) {\rm Tr}(\chi_+)   
 + d_6 \, {\rm Tr} (\bar{B}  \{ \chi_+ , B\} ) {\rm Tr}(\chi_+)
 \nonumber \\   
& + & d_7 \, {\rm Tr} (\bar{B}B) {\rm Tr}(\chi_+) {\rm Tr}(\chi_+)   
 +  d_8 \, {\rm Tr} (\bar{B}B) {\rm Tr}(\chi_+^2) \, \, \, .
\label{leff4}
\end{eqnarray}
We choose to eliminate the $d_6$--term by use of the Cayley--Hamilton
identity Eq.(\ref{CaHa}) for $X = \chi_+ - {\rm Tr}(\chi_+) /3$.
The $d_i$ have dimension mass$^{-3}$. It is important to note that
some of the $d_i$ simply amount to quark mass renormalizations of some
of the dimension two LECs \cite{lelu}. To be specific, one can absorb
the effects of $d_5$ and $d_7$ in $b_F$ and $b_0$, respectively, as follows
\begin{equation}
b_F \to b_F - d_5 \, {\rm Tr}(\chi_+) \, \, ,\quad
b_0 \to b_0 - d_7 \, {\rm Tr}(\chi_+) \, \, .
\label{qmrenor}
\end{equation}
This is a very general phenomenon of CHPT calculations in higher
orders. For example, in $\pi \pi$ scattering there are six LECs at two
loop order $(q^6)$ \cite{bcegs}, but only two new independent terms
$\sim s^3$ and $\sim s \, M_\pi^4$. The other four LECs make the
$q^4$ counter terms $\bar{\ell}_i$ ($i=1,2,3,4)$ quark mass dependent. At
this point, one has two options. One can either treat the higher order
LECs as independent from the lower order ones or lump them together to
mimimize the number of independent terms. In the latter case, one
needs to refit the numerical values of the lower dimension LECs. We
prefer to work with the first option and treat all the $d_i$
separately from the $b_i$.
Therefore, we have fourteen LECs not fixed by chiral symmetry
in addition to  the $F$ and $D$ coupling constants (subject to the
constraint $F+D = g_A =1.25$) from the lowest order Lagrangian
${\cal L}_{\phi B}^{(1)}$.
What we have to calculate are all one--loop
graphs with insertions from ${\cal L}_{\phi B}^{(1,2)}$  and  tree
graphs from ${\cal L}_{\phi B}^{(2,4)}$. We stress that we do not
include the spin--3/2 decuplet in the effective field theory
\cite{dobo}, but rather use these fields to estimate the pertinent
low--energy constants (resonance saturation principle). We therefore
strictly expand in small quark masses and external momenta 
(collectively denoted by $'q \, '$) with no 
recourse to large $N_c$ arguments. 


\section{Baryon masses, $\pi N$ $\sigma$--term  and renormalization}
\label{sec:massigren}

In this section, we assemble the formulae for the  quark
mass squared contribution  to the baryon masses and the
$\sigma$--term. We also discuss the necessary renormalization of the
LECs $d_i$ since at order $q^4$ the meson loop contributions to the
baryon masses and the $\sigma$--terms are no longer finite.

\subsection{Fourth order contribution to the baryon masses}

In general, the quark mass expansion of the octet baryon masses takes the form
\begin{equation}
m = \, \, \krig{m} + \sum_q \, B_q \, m_q + \sum_q \, C_q \, m_q^{3/2} + 
\sum_q \, D_q \, m_q^2  + \ldots
\label{massform}
\end{equation}
modulo logs. The coefficients $B_q, C_q, D_q$ are 
state--dependent. Furthermore, they include contributions proprotional
to the low--energy constants enumerated in the previous section.
Our aim is to evaluate the terms of fourth order
in the chiral counting. The lower order terms $\sim m_q $ and $\sim m_q^{3/2}$ 
for the baryon masses 
are standard, we use here the same notation as Ref.\cite{bkmz}. 
Calculating the one loop graphs shown in Fig.1
and the counter terms $\sim d_i$ not shown in that
figure, the $q^4$ contribution to any octet baryon mass $m_B$ 
can be written as
\beq
m_B^{(4)}  =  \epsilon_{1,B}^P \, M_P^4 + 
\epsilon_{2,B}^{PQ} \, M_P^2 \, M_Q^2 + 
\epsilon_{3,B}^P \, \frac{M_P^4}{\Lambda_\chi^2} \, 
\ln ( \frac{M_P^2}{\lambda^2}) +
\epsilon_{4,B}^{PQ} \, \frac{M_P^2 \, M_Q^2}{\Lambda_\chi^2} \, \ln ( \frac{
  M_P^2}{\lambda^2} )
\, \, ,
\label{mB4}
\eeq
with $P,Q = \pi , K, \eta$ and $\lambda$ the scale of dimensional
regularization. Also, $\Lambda_\chi = 4 \pi F_\pi$ is related to the scale
of chiral symmetry breaking. In fact, this is the canonical prefactor 
appearing in one loop integrals as performed here. We remark that the
graphs with an insertion from ${\cal L}^{(2)}_{\phi B}$ on the baryon 
propagator lead to a state--dependent shift, i.e. they contribute to
the baryon mass splittings at order $q^4$. The exception to this rule 
is the term proportional to $b_0$. Expanded to quadratic order in the
meson fields, it takes the form
\beq
4 \, B \, b_0 \,{\rm Tr}(\bar BB) \, {\rm Tr}({\cal M}) 
- \frac{2}{F^2} \, B \, b_0 \,{\rm Tr}(\bar BB) \, 
{\rm Tr}(\phi^2 \,{\cal M}) + {\cal O}(\phi^4)
\eeq
and thus its contribution to the baryon mass can be completely
absorbed in the octet mass in the chiral limit,
\beq
\krig{m}  \,{\rm Tr}(\bar BB) - b_0 \,\,{\rm Tr}(\bar BB) \, {\rm
  Tr}(\chi_+) = \krig{m}' \,{\rm Tr}(\bar BB) \,\,\, .
\eeq
Therefore, as in  the case of the
$q^3$ calculation \cite{liz,bkmz} one  needs 
additional information (like from one of the $\sigma$--terms) to
disentangle $b_0$ and $\krig m$. 
The explicit form of the state--dependent
prefactors for the nucleon, i.e. the $\epsilon_{i,N}^{P(Q)}$, is
(for the LECs $d_i$  the renormalized, finite values enter as 
denoted by the superscript 'r', see next paragraph, and we do not show
their explicit scale dependence)
\begin{eqnarray}
\epsilon_{1,N}^\pi &=& -4(4d_1^r+2d_5^r+d_7^r+3d_8^r) \, \, , 
\epsilon_{2,N}^{\pi K} = 
8(4d_1^r-2d_2^r-d_5^r-2d_7^r+2d_8^r) \, \, , \nonumber \\
\epsilon_{1,N}^K & = & 16(-d_1^r+d_2^r-d_3^r+d_5^r-d_7^r-d_8^r) 
\, \, , \nonumber \\
\epsilon_{3,N}^\pi &=& 3(b_D + b_F + 2b_0 - b_1 -b_2 -b_3-2b_8) 
- \frac{3(D+F)^2}{4 \krig{m}} \, \, , \nonumber \\
\epsilon_{3,N}^K & = & 2(3b_D - b_F + 4b_0 - 3b_1 +b_2 -3b_3-4b_8) 
- \frac{5D^2 - 6DF +9F^2}{6 \krig{m}} \, \, , \nonumber \\
\epsilon_{3,N}^\eta & = & 2b_0 - 3b_1+ b_2 -\frac{1}{3}b_3-2b_8 
- \frac{(D-3F)^2}{12 \krig{m}} \, \, , 
\epsilon_{4,N}^{\pi \pi}  =   0 
\,\, \, , \, \, \epsilon_{4,N}^{\pi K} = 0
\,\, \, , \nonumber \\
\epsilon_{4,N}^{K \pi} & = & (5D^2-6DF+9F^2) \, (2b_F+b_0) -
\frac{9}{2} (D-F)^2 \, (2b_D+b_0) \nonumber \\ 
& & \qquad \qquad \qquad\qquad \qquad - \frac{1}{6}(D+3F)^2 (3b_0 -2b_D)
\,\, \, , \nonumber \\
\epsilon_{4,N}^{K K} & = & -2(5D^2-6DF+9F^2) \, (-b_D+b_F-b_0) -
9 (D-F)^2 \, b_0 \nonumber \\ 
& & \qquad \qquad \qquad \qquad \qquad - \frac{1}{3}(D+3F)^2 (3b_0 +4b_D)
\,\, \, , \nonumber \\
\epsilon_{4,N}^{\eta \pi} & = & -b_D + \frac{5}{3} b_F
\,\, \, , \quad
\epsilon_{4,N}^{\eta K} =  \frac{8}{3}b_D - \frac{8}{3}b_F 
 \, \, \, \, .
\label{epsnucl}
\end{eqnarray} 
The corresponding coefficients for the $\Lambda$, $\Sigma$ and the
$\Xi$ are collected in Appendix~\ref{app:mbcoeffs}.
A few remarks are in order. First,
we have $\epsilon^\eta_{1,B} =\epsilon^{P \eta}_{2,B} =
\epsilon^{P \eta}_{4,B} = 0$  making use
of the Gell-Mann--Okubo relation for the pseudoscalar meson masses. In
these terms, the deviations from the GMO limit enters only at order
$q^6$. Second, the $\epsilon^{PP}_{4,B}$ could be absorbed in the
$\epsilon^P_{3,B}$. We prefer to keep the more symmetric notation given
in Eq.(\ref{epsnucl}). Third, we stress that the mixed terms 
$\sim M_P^2 \, M_Q^2$ were not considered in most of the existing 
investigations. Notice that we have kept the baryon mass in the chiral
limit. In the fourth order terms, it could simply be substituted by
the physical mass (the difference being of higher order). 

\subsection{Renormalization}

The fourth order contribution to the baryon masses are no longer finite.
The appearance of these divergences is in marked contrast 
with the $q^3$ calculation which is completely finite 
(in the heavy fermion approach). To be precise, we calculate the
baryon self--energy $\Sigma_B (\omega)$, which is related to the
baryon propagator $S_B (\omega)$ via
\begin{equation}
S_B (\omega) =\frac{i}{p\cdot v - \krig{m}- \Sigma_B (\omega)}
\, \, \, ,
\end{equation}
with $p_\mu = \krig{m}v_\mu + l_\mu$, $v \cdot l \ll \krig{m}$ and
$\omega = v \cdot l$. The propagator develops a pole at $p =m_B
\, v$, with $m_B$ the renormalized (physical) baryon mass,
\begin{equation}
m_B = \krig{m} + \Sigma_B (0) \quad . 
\end{equation}
The nucleon wave--function renormalization is determined by the
residue of the propagator at the physical mass,
\begin{equation}
S_B (\omega) =\frac{i\ Z_B}{p\cdot v - m_B} \,\, \, , \quad
Z_B = 1 + \Sigma_B' (0) \, \, \, .
\end{equation}
To calculate the pertinent loop graphs, we use dimensional
regularization 
and all appearing divergences can be absorbed in the LECs $d_i$
\begin{equation}
d_i = d_i^r (\lambda) + \frac{\Gamma_i}{F_\pi^2} \, L
\, \, .
\label{divd}
\end{equation}
with $\lambda$ the scale of dimensional regularization and
\begin{equation}
L = \frac{\lambda^{d-4}}{16 \pi^2} \biggl\lbrace \frac{1}{d-4} -
\frac{1}{2}[\ln (4 \pi) +1 - \gamma_E] \biggr\rbrace
\, \, .
\label{L}
\end{equation}
Here, $\gamma_E = 0.5772215$ is the Euler-Mascheroni constant. The
scale dependence of the $d_i^r (\lambda)$ follows from
Eq.(\ref{divd}):
\begin{equation}
d_i^r (\lambda_2) = d_i^r (\lambda_1) + \frac{\Gamma_i}{(4 \pi F_\pi)^2} 
\ln \frac{\lambda_1}{\lambda_2} \quad .
\end{equation}
In what follows, we set $\lambda =1$~GeV. The corresponding
$\Gamma_i$ read
\begin{eqnarray}
\Gamma_1 &=& -\frac{1}{6}b_1 +\frac{1}{18}b_3
+\frac{1}{36}(7+15D^2+27F^2) b_D \nonumber \\
& + &\frac{1}{72 \krig{m} } (D^2-3F^2)
-\frac{3}{8}\biggl( 
-\frac{1}{2}(D^2 - 3 F^2) b_0 + \frac{2}{3}DFb_F \biggr)
\nonumber \\
\Gamma_2 &=& \frac{1}{4}b_2 +\frac{1}{12}(1 +24D^2)b_F
+\frac{13}{96 \krig{m}} DF
\nonumber \\
\Gamma_3 &=& -\frac{3}{4}b_1 -\frac{1}{12}b_3
+\frac{1}{4}(2+3(D^2-3F^2))b_D \nonumber \\ 
& + &\frac{7}{128 \krig{m}} (D^2-3F^2)
- \frac{3}{4}\biggl( 
-\frac{3}{4}(D^2-3F^2)  b_0 - 3 D F b_F \biggr)
\nonumber \\
\Gamma_4 &=& \frac{3}{2}b_1 -\frac{1}{18}b_3
+\frac{1}{2}\biggl( -\frac{11}{9}-\frac{9}{4}\bigl(\frac{68}{27}D^2
-4F^2 \bigr) \biggr) b_D  \nonumber \\
& - &\frac{1}{8 \krig{m} } \bigl( D^2- 3 F^2 \bigr)
- \frac{9}{4}\biggl( 
\frac{1}{3}D^2 -F^2   \biggr)  b_0  \\
\Gamma_5 &=& -\frac{13}{18}b_2 +\frac{1}{2}(\frac{11}{9} 
- \frac{13}{3}D^2 - 3F^2)b_F -\frac{5}{2} D F b_0
-\frac{13}{36 \krig{m}} DF
\nonumber \\
\Gamma_7 &=& -\frac{1}{4}b_1 -\frac{35}{108}b_3
-\frac{11}{18}b_8 -\frac{7}{128 \krig{m}} 
(\frac{35}{27}D^2+F^2)
\nonumber \\
&-&\frac{1}{8}\biggl( -\frac{22}{9}+\frac{34}{3}D^2+6F^2 \biggr) b_D
 + \frac{1}{8}\biggl(\frac{44}{9}-\frac{43}{3}D^2- 21 F^2
\biggr)b_0    \nonumber \\
\Gamma_8 &=& -\frac{1}{4}b_1 -\frac{17}{36}b_3
-\frac{5}{6}b_8 -\frac{7}{128 \krig{m}} 
(\frac{17}{9}D^2+F^2) -3 D F b_F
\nonumber \\
&-&\frac{1}{8}\biggl( -\frac{14}{9}+\frac{2}{3}D^2+6F^2 \biggr)b_D
+ \frac{1}{2}\biggl(\frac{5}{3}-\frac{3}{4}(D^2-3F^2) \biggr)  b_0 
 \nonumber  
\end{eqnarray}    
We note that for performing this renormalization, one has to include
the terms 
\begin{eqnarray} 
& &-2i \biggl(\frac{10}{3}D^2+F^2\biggr) \, {\rm Tr}(\bar{B} [v \cdot D, [v
\cdot D, [ v\cdot D,B]]])  \nonumber \\ & & + i\frac{3}{4} (D^2 -3F^2) 
\, {\rm Tr}(\bar{B} \{ \chi_+, [ v \cdot D, B]\})
 - i \frac{5}{2} DF \, {\rm Tr}(\bar{B} [ \chi_+, [ v \cdot D, B]])
\nonumber \\
& & -i \frac{3}{2} (\frac{13}{9}D^2 +F^2) \, {\rm Tr}(\bar{B}  [ v \cdot D, B])
{\rm Tr}( \chi_+)  \nonumber \\
& & - \frac{3}{8\krig{m}}(D^2-3F^2)
{\rm Tr}(\bar{B} \{ \chi_+ , [D^\mu, [ D_\mu,B]]\} ) \no \\
& & + \frac{5}{4 \krig{m}}DF
{\rm Tr}(\bar{B} [ \chi_+ , [D^\mu, [ D_\mu,B]]] ) \no \\
& & + \frac{3}{4}(\frac{13}{9}D^2+F^2)
{\rm Tr}(\bar{B}  [D^\mu, [ D_\mu,B]] ) {\rm Tr}(\chi_+) \no \\
& & + \biggl\{  \frac{3}{\krig{m}}(D^2-3F^2)  -3 \biggl(
 \frac{1}{3}D^2 \, b_0  -b_D(D^2 -3F^2) - 6DFb_F \biggr)
\biggr\} \nonumber \\ & & \qquad \qquad \qquad\qquad \qquad \qquad 
\qquad \qquad \qquad
\times {\rm Tr}(\bar{B} \{ \chi_+ , [v \cdot D, [ v\cdot D,B]]\} ) 
\nonumber \\
& & + \biggl\{ -\frac{10}{\krig{m}}DF - 4 \biggl(
b_F \biggl(5D^2+9F^2\biggr) -10 DF\,b_D \biggr)
\biggr\} \, {\rm Tr}(\bar{B} [ \chi_+ , [v \cdot D, [ v\cdot D,B]]] )
\nonumber \\  
& & + \biggl\{-\frac{6}{\krig{m}}\bigl(\frac{13}{9}D^2+F^2\bigr) 
-6 \biggl(
b_0 \biggl(\frac{8}{3}D^2+6F^2\biggr) + b_D \biggl(2 F^2
+\frac{26}{9}D^2 \biggr) + 4DFb_F \biggr)
\biggr\} \nonumber \\ & & \qquad \qquad \qquad\qquad \qquad \qquad 
\qquad \qquad \qquad
\times {\rm Tr}(\bar{B}  [v \cdot D, [ v\cdot D,B]] ) {\rm Tr}(\chi_+) 
\nonumber \\ 
& & + \frac{1}{\krig m} (10D^2+18F^2) \, {\rm Tr}(\bar{B} [v \cdot D, [v
\cdot D, [ D^\mu, [  D_\mu,B]]]]) \no \\
& & -\frac{10}{\krig m}\biggl(\frac{10}{3}D^2+6F^2\biggr)
\, {\rm Tr}(\bar{B} [v \cdot D, [v
\cdot D, [ v\cdot D, [ v \cdot  D,B]]]])
\label{leffrenor}
\end{eqnarray}
in the effective Lagrangian (an overall factor $L F_\pi^{-2}$ has been
scaled out and the finite pieces related to these operators can be neglected). 
These, however, do not directly contribute to the masses
and $\sigma$--terms and are therefore not listed in section~3.1. It is
obvious that the dimension two terms also enter the renormalization
procedure and thus the finite constants $b_i$ appear in the
$\Gamma_i$. This is a general feature of any renormalization beyond
${\cal O}(q^3)$ in heavy baryon CHPT.

\subsection{Pion/kaon--nucleon $\sigma$--terms and strangeness content}
\label{sec:sigmaterms}

Further information on the scalar sector of baryon CHPT is given by
the scalar form factors or $\sigma$--terms which measure the strength
of the various matrix elements $m_q \bar{q}q$ $(q=u,d,s$) in the
proton. In a mass--independent renormalization scheme, one can define
the following renormalization--group invariant quantities:
\begin{eqnarray}
\label{defsigma}
\sigma_{\pi N} (t) & = & \hat m \, <p' \, | \bar u u + \bar d d| \, p>
\, \, \, , \nonumber \\
\sigma_{KN}^{(1)} (t) & = & \frac{1}{2}(\hat m + m_s) \, 
<p' \, | \bar u u + \bar s s| \, p> \, \, \, , \nonumber \\
\sigma_{KN}^{(2)} (t) & = & \frac{1}{2}(\hat m + m_s) \, 
<p' \, | -\bar u u + 2\bar d d + \bar s s| \, p> \, \, \, , 
\end{eqnarray}
with $t = (p'-p)^2$ the invariant momentum transfer squared.
The explicit form of the $t$-dependent $\sigma$--terms is rather lengthy
and not very instructive. Here, we just discuss some general aspects
and refer the reader to Ref.\cite{bora} for details. The most striking
new feature compared to the ${\cal O}(q^3)$ analysis
\cite{bkmz,bkkm,gss} is the appearance of $t$--dependent divergences. To
be more specific, consider a typical diagram as shown in Fig.2.
It is most convenient to calculate these diagrams in the Breit--frame 
\cite{bkkm}. For the case at hand, we get
\begin{eqnarray}
& &I_\sigma (t)  =  \int \frac{d^4k}{(2 \pi)^4} \frac{i^2 \, k \,
  (k+l)}{(k^2-M_a^2)((k+l)^2-M_a^2)} = i \biggl\{ (4M_a^2-t) \, L
 \\  &  & \!\!\!\!\!\!\!\!\!\!\!\!\! 
- \frac{1}{16\pi^2} \biggl( M_a^2 - \frac{t}{2} -\frac{1}{2}(4M_a^2-t)
\ln\frac{M_a^2}{\lambda^2} - (2M_a^2-t)
\frac{\sqrt{4M_a^2-t}}{\sqrt{t}}\arcsin \bigl( \frac{\sqrt{t}}{2M_a}\bigr)
 \biggr) \biggr\} \nonumber
\label{Isigt}
\end{eqnarray} 
where we have suppressed the pertinent Clebsch--Gordan coefficient and
$M_a$ stands for a Goldstone boson mass in the SU(3) flavor basis.
The renormalization of this diagram and the other contributing to the
$\sigma$--terms is somewhat tricky and is described in
 appendix~\ref{app:renorm}. Note that when combining all the terms 
from the various Feynman graphs (after renormalization), 
one is indeed left with terms proportional to const$\cdot t$. In
fact, these $t$--dependent coefficients can not easily be estimated. Also,
they are different for $\sigma_{\pi N} (t)$
and $\sigma^{(1,2)}_{KN} (t)$, respectively.  The empirical
information on $\sigma_{\pi N} (2M_\pi^2) - \sigma_{\pi N} (0) = 15 \,
$MeV \cite{gls} is thus not sufficient to make predictions for the two
$KN$ $\sigma$-terms shifts, $\sigma_{KN}^{(1,2)} (2M_K^2) 
- \sigma_{KN}^{(1,2)} (0)$. We will therefore only give some rough
estimates by simply parametrizing these t--dependent pieces in terms
of coefficients which we can estimate to be of order one. Note that
to order $q^3$, the shifts from $t=0$ to the respective Chang--Dashen
points are finite and free of any LEC.  In view of this, the nice
result found in Ref.\cite{bkmz} for $\Delta \sigma_{\pi N} \simeq 15
\,$MeV by including the $\Delta(1232)$ explicitely as intermediate 
state must be considered accidental.

At $t=0$, the corresponding formulae simplify considerably. The fourth
order contribution to $\sigma_{\pi N}(0)$ takes the form
\begin{equation}
\sigma_{\pi N}^{(4)}(0)  = 
M_\pi^2 \, \biggl[ \, \epsilon_{1,\sigma}^P \, M_P^2 + 
\epsilon_{2,\sigma}^P \, 
\frac{M_P^2}{\Lambda_\chi^2} \, \ln(\frac{M_P^2}{\lambda^2}) +
\epsilon_{3,\sigma}^{PQ} \, \frac{M_P^2}{\Lambda_\chi^2} 
\, \ln(\frac{M_Q^2}{\lambda^2}) \, \biggr]
\, \, ,
\label{sigma4}
\end{equation}
with the $\epsilon_{i,\sigma}^{P(Q)}$ are related to the 
$\epsilon_{i,N}^{P(Q)}$, Eq.(\ref{epsnucl}), and they are collected in 
Appendix~\ref{app:sigmaeps}. These relations are, of course, a
consequence of the Feynman--Hellmann theorem which states that
\begin{equation}
\sigma_{\pi N} (0)  = \hat m \, \frac{\partial \,m_N}{\partial \, \hat
  m} \quad . 
\label{FeHe}
\end{equation}
The derivatives with respect to  the quark masses can be converted
into derivatives with respect to the Goldstone boson masses squared,
for example
\begin{equation}
\hat m \frac{\partial}{\partial \hat m} \, M_K^4 \, \ln 
\frac{M_K^2}{\lambda^2} = M_\pi^2 \,  M_K^2 \,  \biggl\{
\frac{1}{2} + \ln \frac{M_K^2}{\lambda^2} \biggr\} \, \, \, .
\end{equation}
Eq.(\ref{FeHe}) can therefore be evaluated easily.
By a similar reasoning, one can deduce the strangeness content of the
proton, i.e the strength of the matrix element $<p|m_s \bar  s s|p>$,
\begin{equation}
<p|m_s \bar  s s|p> = m_s \, \frac{\partial \,m_N}{\partial \,m_s}
\quad , 
\label{defmsssbar}
\end{equation}
and  the strangeness fraction $y$,
\begin{equation}
y = \frac{ 2 \, <p| \bar  s s|p>}{<p|\bar u u + \bar d d |p>} =
\frac{M_\pi^2}{\sigma_{\pi N} (0)} \biggl( M_K^2 -
\frac{1}{2} \, M_\pi^2 \biggr)^{-1}   \, 
 m_s \, \frac{\partial \,m_N}{\partial \,m_s}
\quad .
\label{defy}
\end{equation}
The fourth order contribution to $y$, denoted by $y^{(4)}$, can be
decomposed in the standard fashion (for the lower orders, see
\cite{bkmz})
\begin{equation}
y^{(4)}  = \frac{2 \,M_\pi^2}{\sigma_{\pi N} (0)} \, 
\biggl[ \, \epsilon_{1,y}^P \, M_P^2 + 
\epsilon_{2,y}^P \, 
\frac{M_P^2}{\Lambda_\chi^2} \, \ln(\frac{M_P^2}{\lambda^2}) +
\epsilon_{3,y}^{PQ} \, \frac{M_P^2}{\Lambda_\chi^2} 
\, \ln(\frac{M_Q^2}{\lambda^2}) \, \biggr]
\, \, ,
\label{y4}
\end{equation}
and the $\epsilon_{i,y}^{P(Q)}$ are again related to the
$\epsilon_{i,N}^{P(Q)}$, see appendix~\ref{app:sigmaeps}. The more
lengthy expressions for the two $KN$ $\sigma$--terms can be found in
Ref.\cite{bora}. This completes the formalism to study the scalar
sector of baryon CHPT to fourth order in the meson masses.


\section{Low--energy constants from resonance saturation}
\label{sec:LECS}

Clearly, we are not able to fix all the low--energy
constants appearing in ${\cal L}_{\phi B}^{(2,4)}$
from data, even if we would resort to large $N_c$ arguments.
We will therefore use the principle of resonance saturation to
estimate these constants. This works very accurately in the meson
sector \cite{reso,reso1,reso2}. In the baryon case, one has to account for 
excitations of meson ($R$) and baryon ($N^*$) resonances. One writes
down the effective Lagrangian with these resonances chirally coupled to
the Goldstone bosons and the baryon octet, calculates the  Feynman 
diagrams pertinent to the process under consideration and, finally,
lets the resonance masses go to infinity (with fixed ratios of
coupling constants to masses). This generates higher order terms in
the effective meson--baryon Lagrangian with coefficients expressed in
terms of a few known resonance parameters. Symbolically, we can write
\begin{equation}
\tilde{{\cal L}}_{\rm eff} [\, U,B,R,N^* \, ] \to 
{\cal L}_{\rm eff} [\, U,B \, ]  \, \, . 
\end{equation}
It is important to stress that only after integrating out the heavy
degrees of freedom from the effective field theory, one is allowed to
perform the heavy mass limit for the ground--state baryon octet.
Here, there are two relevant types of contributions. One comes from 
the excitation of intermediate baryon resonances, in particular
from the spin-3/2 decuplet states. Concerning the higher baryon resonances,
only the parity--even spin--1/2 octet including the Roper $N^* (1440)$
plays some role. In addition, there is $t$--channel scalar
and vector meson exchange. It was already shown in Ref.\cite{bmlet}
that the vectors do not come in at order $q^4$. It is important to
stress that for the resonance contribution to the baryon masses, one
has to involve Goldstone boson loops. This is different from the
normal situation like e.g. in form factors or scattering processes.
One could argue that scalar meson exchange alone should provide the
necessary strength of the scalar--isoscalar operators
$\sim$Tr$(\chi_+)$. However, taking the phenomenological bounds on the
scalar masses and coupling constants, these contributions are too
small to explain e.g. the strengths of the symmetry breakers
$b_{0,D,F}$ which have been determined to order $q^3$ in \cite{bkmz}.
To be specific, if one performs the calculation in SU(2) and demands
that the LEC $c_1$, which can be fixed from the $\pi N$ $\sigma$--term,
to be saturated by scalar meson exchange only, one would need a mass to 
coupling constant ratio of $M_s / g_s= 220$~MeV. Here, $g_s$ refers to
the SU(2) coupling $g_s \bar{\Psi}_N \Psi_N \, S$.
Such a number can not be obtained from standard scalar
meson masses and their couplings to the nucleon (baryons). In
Ref.\cite{bkmz} it was, however, shown, that one--loop graph with
intermediate decuplet states can effectively produce such couplings.
However, treating the resonances relativistically leads to four
major complications. 
\begin{enumerate}

\item[$\circ$] First, terms arise which are non--analytic in the
meson masses. Clearly, to avoid any double counting and 
to be consistent with the
requirements from analyticity, one should only consider the analytic
terms in the meson masses generated by such loop diagrams. In fact, standard
tree level resonance excitation generates the dimension two operators
$b_{1,\ldots,8}$ in the conventional fashion. Calculating tadpole
diagrams to order $q^4$ with these $b_i$ generated from resonance
exchange exactly produces the non--analytic terms 
$\sim M_\phi^4 \ln M_\phi^2$ which also arise from the one--loop
graphs with intermediate decuplet states. This is why we have to
dismiss such non--analytic contributions in the Goldstone boson masses
in our estimation of the scalar--isoscalar LECs from loop graphs. 
We remark that this kind of equivalence has been already discussed for
the SU(2) LEC $c_2$ which enter the calculation of the nucleon
electromagnetic polarizabilities to order $q^4$, see \cite{bkms}
\cite{butnat}.

\item[$\circ$] Second, to the order we are working, even the remaining analytic
pieces are divergent (this was incorrectly treated in
ref.\cite{bmlet}). Therefore, we can only determine the analytic
resonance contribution up to three renormalization constants, two
related to the decuplet ($\beta_\Delta$, $\delta_\Delta$) 
and one to the Roper--octet ($\beta_R$). We solve this
problem in the following manner. In addition to these three constants,
we have as further unknowns the scalar couplings $F_S$ and $D_S$ and
the octet baryon mass in the chiral limit. Fitting the four octet
baryon masses and $\sigma_{\pi N} (0)$ allows us to express all LECs
and physical observables in terms of $\beta_R$ solely. Fortunately,
all quantities are very insensitive to the choice of this parameter
which we can bound by some phenomenological  argument.

\item[$\circ$] Third, since the baryon excitations are treated
  relativistically as explained
above, there is no more strict power counting \cite{gss} and thus we
must address the question of two--loop contributions from diagrams 
with baryon excitations. To leading order in the resonance mass
expansion, they give rise to the same
divergences as the one loop graphs  and the corresponding corrections can
be completely absorbed in a redefinition of the parameters appearing at
one loop. We also show that there is an additional divergence at order
$M_\phi^4$ in the Roper--octet graphs which
modifies the finite one--loop parameter $\delta_R$. However, the contributions
to the baryon masses and the $\sigma$--terms $\sim \delta_R$ are very
small and thus one does not need to know this parameter accurately.
Using dimensional analysis, the two--loop corrections to $\beta_R$ are
found to be modest. For details, see paragraph~\ref{sec:twoloop}.

\item[$\circ$] Fourth, the renormalization of the $\sigma$--terms is
somewhat complicated. The divergences of the scalar form factors are
renormalized through terms of the form 
$\partial {\cal L} / \partial {m_q}$. In general, one has
$\partial {\cal L} / \partial {m_u} \neq 
 \partial {\cal L} / \partial {m_d} \neq
 \partial {\cal L} / \partial {m_s}$. If one calculates in the isospin
 limit $m_u = m_d$, one can not disentangle the derivatives with respect
 to $m_u$ and $m_d$ any more. For the general renormalization
 procedure, one has to work with $m_u \neq m_d$. Applying resonance
 saturation, one effectively determines the LECs of the Lagrangian but
 not the ones related to the operators $\bar u u$, $ \bar d d$ and
 $\bar s s$. By use of the Feynman--Hellmann theorem, the LECs related
 to the operators $\bar q q$ can be fixed from the ones which appear
 in the calculation of the baryon masses. This general theorem is
 proven in ref.\cite{bora}. It means that here we are only able to 
 give the full order $q^4$ contribution to the pion--nucleon
 $\sigma$--term at $t=0$ and can only determine $\sigma_{KN}^{(1,2)} (0)$ up
 some renormalization constant (since we work in the isospin limit).
 We also point out that we have not
 yet been able to generalize these methods to the momentum--dependent
 parts of the scalar form factors since there integrals appear which
 we can not work out analytically (for details, see ref.\cite{bora}). 
\end{enumerate}

Let us now consider the one--loop contributions from the 
decuplet, the $\frac{1}{2}^+$--octet and from the scalar mesons, in order.

\subsection{Decuplet contribution to the low--energy constants}

Consider first the decuplet contribution. We treat these field
relativistically and only at the last stage let the mass become very
large. The pertinent interaction Lagrangian between the spin--3/2
fields (denoted by $\Delta$), the baryon octet and the Goldstone bosons reads
\begin{equation}
{\cal L}_{\Delta B \phi} = 
\frac{{\cal C}}{2} \, \biggl\{ \bar{\Delta}^{\mu ,abc} \, 
 \Theta_{\mu \nu} (Z) \, (u^\nu)_a^i \, B_b^j \, 
\epsilon_{cij}+ \bar{B}^b_j \, (u^\nu)^a_i \,  \Theta_{\nu \mu} (Z) \,
{\Delta}^\mu_{abc} \,\epsilon^{cij}\, \biggr\} \, \,  \, ,
\label{lmbd}
\end{equation}
with $'a, b, \ldots , j'$ are SU(3)$_f$ indices
and the coupling constant ${\cal C} =1.2 \ldots 1.8$ can be
determined from the decays $\Delta \to
B \pi$. The Dirac matrix operator $\Theta_{\mu \nu} (Z)$ is given by
\begin{equation}
\Theta_{\mu \nu} (Z) = g_{\mu \nu} - \biggl(Z + \frac{1}{2} \biggr) \,
\gamma_\mu \, \gamma_\nu \, \, \, \, .
\label{theta}
\end{equation}
For the off--shell parameter $Z$, we use $Z = -0.3$ from the
determination of the $\Delta$ contribution to the $\pi N$ scattering
volume $a_{33}$ \cite{armin}. That is also consistent with recent
studies of $\Delta (1232)$ contributions to the nucleons'
electromagnetic polarizabilities \cite{bkmzas} and to threshold 
pion photo-- and electroproduction \cite{bkmpe}. 
For the  processes to be discussed,
we only need the expanded form of $u_\mu$,
\begin{equation}
(u_\mu)^i_a = -\frac{1}{F_\pi} \partial_\mu \, \phi_a^i + 
{\cal O}(\phi^2) \quad .
\end{equation}
The propagator of the spin--3/2 fields has the form
\begin{equation}
G_{\beta \delta} (p) = -i\frac{p \!\!/ + m_\Delta}{p^2-m_\Delta^2} 
\, \biggl( g_{\beta \delta} - \frac{1}{3} \gamma_\beta \gamma_\delta -
\frac{2 p_\beta p_\delta}{3m_\Delta^2} + \frac{p_\beta \gamma_\delta -
p_\delta \gamma_\beta}{3 m_\Delta} \, \biggr) \, \, \, ,
\end{equation}
with $m_\Delta = 1.38$~GeV  the average decuplet mass.
We have to evaluate diagrams as shown in Fig.~3. With the labeling of
the momenta as in that figure, this leads to
\begin{equation}
I_\Delta (p) = \frac{-i \,{\cal C}^2}{2F^2_\pi} \int \frac{d^4
k}{(2\pi)^4} \frac{(k-p)^\sigma \,
\Theta_{\sigma \rho} (Z) \, G^{\rho \mu} (k) \, \Theta_{\mu \nu} (Z)
\, (k-p)^\nu }{(k-p)^2-M_a^2+i\epsilon}\, \, \, .
\label{Idelta}
\end{equation}
where the pertinent Clebsch--Gordan coefficient has been omitted and $M_a$
as defined after Eq.(\ref{Isigt}). 
This integral is evaluated on the mass--shell of the
external baryons, i.e at $p \!\!/ = \krig{m}$, and
split into various contributions according to the power of momenta in
the numerator and the number of propagators. Each such term is then is
expanded in powers of Goldstone boson masses up-to-and-including
${\cal O}(M_a^4)$ (modulo logs).
Only then the large mass limit of the decuplet is
taken. This then gives the contribution to the various LECs $b_i$ and
$d_i$ modulo some renormalization constants. To be specific, consider 
the scalar integral to which all contributions arising in
evaluating $I_\Delta$ can be reduced
\begin{eqnarray}
& &i \, \int \frac{d^4 k}{(2\pi)^4} \frac{1 }
{[(k-p)^2-M_a^2+i\epsilon][k^2 -m^2_\Delta+ i\epsilon]} \nonumber
\\ 
&=& \biggl\{2L + \frac{1}{16\pi^2} \biggl[-1 +
\frac{m_\Delta^2}{\krig{m}^2} \ln({m_\Delta^2 \over \lambda^2})
- \frac{1}{\krig{m}^2}(m_\Delta^2-\krig{m}^2) \ln({m_\Delta^2
-\krig{m}^2 \over \lambda^2}) \nonumber \\
&+& \frac{1}{\krig{m}^2 (m_\Delta^2-\krig{m}^2)} \biggl( \krig{m}^2
-m_\Delta^2 \ln({m_\Delta^2 \over \lambda^2}) 
+ (m_\Delta^2+\krig{m}^2) \ln({m_\Delta^2
-\krig{m}^2 \over \lambda^2}) \biggr) \, M_a^2 \nonumber \\
&+& \frac{1}{2 (m_\Delta^2-\krig{m}^2)^3} \biggl( -\krig{m}^2
-2m_\Delta^2 \ln({m_\Delta^2 \over \lambda^2}) + 4m_\Delta^2 \ln({m_\Delta^2
-\krig{m}^2 \over \lambda^2}) \biggr) \, M_a^4 \nonumber \\
&-& \frac{1}{m_\Delta^2-\krig{m}^2} M_a^2 \ln ({M_a^2 \over \lambda^2})
- \frac{m_\Delta^2}{(m_\Delta^2-\krig{m}^2)^3} M_a^4 \ln ({M_a^2 \over
  \lambda^2})  \biggr] \biggr\} \,\, .
\label{Idbsp}
\end{eqnarray}
In total, the $M_a^2 \ln M_a^2$ terms cancel in $I_\Delta$.  
As explained before, the remaining non--analytic pieces $\sim M_a^4 \ln
M_a^2$ appearing in these
integrals should not be retained, see also below. We can now work out
the complete integral $I_\Delta (p \!\!\!/= \krig{m})$ 
and find at the scale $\lambda = m_\Delta$
\begin{equation}
\!\!\!\!\!\!\!\!\!\!\!\!
I_\Delta^{\rm analytic}  = \frac{-i \,{\cal
 C}^2}{6 F_\pi^2} \, \biggl\{ -(2 Z +1 ) m_\Delta
\, M_a^2  -  (4Z^2+5Z-\frac{3}{2})  
\frac{1}{m_\Delta}  \,M_a^4 \biggr\} \, 2L + \ldots
\label{Ideltaresz}
\end{equation}
where the ellipsis stands for constant and divergent pieces which
have to be taken care of by standard mass renormalization terms of the
type $\delta m_\Delta \, {\rm Tr}(\bar{\Delta} \Delta)$, 
see e.g. Refs.\cite{bkmz,gss}, and for subleading terms in the
$1/m_\Delta$--expansion at orders $M_a^2$ and $M_a^4$. If one were
to use the subleading finite pieces to estimate the LECs, the
resulting values would be much too small.
One notices  that in this relativistic
treatment, the dimension two and four LECs
are no longer finite (as in the heavy baryon approach) \cite{gss}. 
The structure of Eq.(\ref{Ideltaresz}) shows that the first and second 
term in the curly brackets will contribute to the $b_{0,D,F}$,
and the $d_i$, respectively, after renormalization 
\begin{equation}
I_\Delta^{\rm analytic}  = \frac{-i \,{\cal C}^2}{2 F_\pi^2} \, 
\frac{1}{16\pi^2} \, \biggl\{ - m_\Delta \, M_a^2 \, \alpha_\Delta^{(2)}
+\frac{1}{m_\Delta}  \,M_a^4 \, \alpha_\Delta^{(4)} \biggr\} \,\, ,
\label{Ideltares}
\end{equation}
where $\alpha_\Delta^{(2,4)}$ are the finite renormalization constants
related to the divergences appearing at order $M_a^2$ and $M_a^4$, 
respectively. The $b_{1,2,3,8}$ are calculated from
standard pion--nucleon scattering tree graphs with intermediate
decuplet states (for details, see ref.\cite{bora}). 
As an important check, feeding these
back into the tadpole diagrams one finds non--analytic pieces which
exactly agree with the ones dropped from the integral
$I_\Delta$. These have the generic form
\beq
I_\Delta^{\rm non-analytic} = (2Z^2+Z-1) \frac{M_a^4}{m_\Delta} \ln
\frac{M_a^2}{\lambda^2} \,\,\, . 
\eeq
This proves the consistency of our procedure. Evaluating now the
pertinent Clebsch--Gordan coefficients and defining
\begin{equation}
\epsilon_\Delta = \frac{{\cal C}^2}{12 m_\Delta} (2Z^2 + Z -1)\, \, ,
\, \, 
\beta_\Delta = \frac{{\cal C}^2}{128 \pi^2 F_\pi^2} 
\, m_\Delta \, \alpha_\Delta^{(2)} \, \, , \, \, 
\delta_\Delta = -\frac{{\cal C}^2}{256 \pi^2 F_\pi^2} 
\, \frac{1}{m_\Delta} \, \alpha_\Delta^{(4)} \, \, ,
\label{ebddel}
\end{equation}
we arrive at the decuplet contribution to the various LECs
\begin{eqnarray}
& & b_0^\Delta  = \frac{7}{3}\, \beta_\Delta \, \, , \, \,  
    b_D^\Delta  = -\beta_\Delta \, \, , \, \,  
    b_F^\Delta  = \frac{5}{6}\, \beta_\Delta \, \, , \nonumber \\  
& & b_1^\Delta  = -\frac{7}{12}\, \epsilon_\Delta \, \, , \, \,  
    b_2^\Delta  = \epsilon_\Delta \, \, , \, \,  
    b_3^\Delta  = -\frac{3}{4}\, \epsilon_\Delta \, \, , \, \,  
    b_8^\Delta  = \frac{3}{2}\, \epsilon_\Delta \, \, , \nonumber \\  
& & d_1^\Delta  = -\frac{1}{18}\, \delta_\Delta \, \, , \, \,
    d_2^\Delta  = -\frac{1}{4}\, \delta_\Delta \, \, , \, \,
    d_3^\Delta  = -\frac{1}{2}\, \delta_\Delta \, \, , \, \,
    d_4^\Delta  =  \frac{5}{9}\, \delta_\Delta \, \, , \nonumber \\
& & d_5^\Delta  =  \frac{13}{36}\, \delta_\Delta \, \, , \, \,
    d_7^\Delta  =  \frac{19}{144}\, \delta_\Delta \, \, , \, \,
    d_8^\Delta  =  \frac{3}{4}\, \delta_\Delta \, \, ,
\label{lecsdel}
\end{eqnarray}
where $\beta_\Delta$ and $\delta_\Delta$ have to be determined as
described below.

\subsection{1/2$^+$--octet contribution to the low--energy constants}

The next multiplet of excited states is the octet of 
even--parity Roper--like spin--1/2
fields. While it was argued in Ref.\cite{dobo} that these play no
role, a more recent study seems to indicate that one can not
completely neglect contributions from these states to e.g. the
decuplet magnetic moments \cite{banmi2}. It is thus important to
investigate the possible contribution of these baryon resonances to
the LECs. The octet consists of the $N^* (1440)$, the $\Sigma^*
(1660)$, the $\Lambda^* (1600)$ and the $\Xi^*$. For the mass of the
latter, we use the Gell-Mann--Okubo formula,
\begin{equation}
m_{\Xi^*} = \frac{3}{2} \, m_{\Lambda^*} +\frac{1}{2} \, m_{\Sigma^*}
- m_{N^*} = 1.79 \, \, {\rm GeV} \quad ,
\end{equation}
so that we find for the average Roper octet mass 
\begin{equation}
 m_{R} = 1.63 \, \, {\rm GeV} \quad ,
\end{equation}
which is 490 MeV above the mean of the ground state octet. Notice that we
denote the spin--1/2$^+$ octet by '$R$'.\footnote{This should not be
  confused with the same symbol denoting any meson resonance.} The
effective Lagrangian of the Roper octet coupled to the ground state
baryons takes the form
\begin{eqnarray}
{\cal L}_{\rm eff} (B,R) & = & {\cal L}_0 (R) + {\cal L}_{\rm int} (B,R)
\, \, , \nonumber \\
{\cal L}_0 (R) & = & i \, {\rm Tr}( \bar{R} \gamma^\mu [D_\mu , R])
- m_R \, {\rm Tr}(\bar{R}R) \, \, , \nonumber \\  
{\cal L}_{\rm int} (B,R) & = & \frac{D_R}{4} \, \biggl[ 
{\rm Tr}( \bar{R} \gamma^\mu \gamma_5 \{u_\mu , B \} ) + {\rm h.c.} \biggr]
\nonumber \\ & + & \frac{F_R}{4} \, \biggl[ 
{\rm Tr}( \bar{R} \gamma^\mu \gamma_5 [u_\mu , B ] ) + {\rm h.c.} \biggr]
\, \, .
\label{leffrop}
\end{eqnarray}
The numerical values of the axial--vector coupling constants $D_R$ and
$F_R$ are determined in appendix~\ref{app:pararoper} (generalizing the
results for the two--flavor case \cite{bkmpipin}). 
With these definitions, we evaluate the same type of graph as shown in
Fig.~3 (dropping again the Clebsch--Gordan coefficients),
\begin{equation}
I_R (p) = - \int \frac{d^4 k}{(2\pi)^4} \frac{(p\!\!/ - k \!\!/) \,
(k \!\!/ - m_R) \, (p\!\!/ - k \!\!/) \,}
{[(p-k)^2-M_a^2+i\epsilon][k^2 - m_R^2 + i\epsilon]}\, \, \, .
\label{Iroper}
\end{equation}
which by inspection leads to the same kind of terms as in
Eq.({\ref{Idelta}), i.e. with $k^n$ ($n=0,1,2,3$) in the numerator
combined with one or two propagators.  Expanding these terms again 
first in powers of the 
Goldstone boson mass $M_a$ and then in powers of $1/m_R$, one finds
at the scale $\lambda = m_R$ that the terms of order $M_a^2$ diverge
whereas the ones of order $M_a^4$ are finite. After appropriate 
renormalization, we have
\begin{equation}
I_R^{\rm analytic} (p \!\!/  =  \krig{m}) = -\frac{i}{16 \pi^2} \, \biggl\{
m_R \, M_a^2 \, \alpha_R^{(2)} + \frac{\krig{m}}{2 m_R^2} \, M_a^4
\biggr\} \, \, \, . 
\label{Iroperres}
\end{equation}
As before, the contributions to the $b_{1,2,3,8}$ are calculated from
tree--level graphs with intermediate Roper states (for details, 
see ref.\cite{bora}). 
Evaluating now the pertinent Clebsch--Gordan coefficients and defining
\begin{equation}
\epsilon_R = -\frac{1}{8 m_R} \, \, , \, \, 
\beta_R = \frac{1}{256 \pi^2 F_\pi^2} \, m_R \alpha_R^{(2)} \, \, \, , \, \, 
\delta_R = \frac{1}{1024 \pi^2 F_\pi^2} \, \frac{\krig{m}}{m_R^2}  
\, \, \, ,  
\label{ebdrop}
\end{equation}
we arrive at spin--1/2$^+$ octet contribution to the various LECs
\begin{eqnarray}
& & b_0^R  = (\frac{13}{9}D_R^2 + F_R^2)\, \beta_R \, \, , \, \,  
    b_D^R  = \frac{1}{2} (3F_R^2-D_R^2) \, \beta_R \, \, , \, \,  
    b_F^R  = \frac{5}{3}\, D_R \, F_R \beta_R \, \, , \nonumber \\  
& & b_1^R  = - (D_R^2-3F_R^2) \, \frac{\epsilon_R}{6} \, \, , \, \,  
    b_2^R   = D_R \, F_R \, \epsilon_R \, \, , \, \,  
    b_3^R  = 0 \, \, , \, \,  
    b_8^R  = \frac{1}{3}\, D_R^2 \, \epsilon_R \, \, , \nonumber \\  
& & d_1^R  = - (D_R^2 - 3F_R^2)\, \frac{\delta_R}{18} \, \, , \, \,
    d_2^R  = -\frac{1}{2}\, D_R \, F_R \, \delta_R \, \, , \, \,
    d_3^R  = - (D_R^2 -3 F_R^2)\, \frac{\delta_R}{4} \, \, ,
    \nonumber \\
& & d_4^R  =  (D_R^2 -3 F_R^2)\, \frac{\delta_R}{3} \, \, , \, \,
    d_5^R  =   D_R \, F_R \, \frac{13 \delta_R}{18} \, \, , \, \,
    d_7^R  =  (\frac{35}{26}D_R^2 + F_R^2) \, \frac{\delta_R}{16}
               \, \, , \nonumber \\
& & d_8^R  =  \frac{1}{4}\, (\frac{17}{9}D_R^2 + F_R^2) \, \delta_R \, \, ,
\label{lecsrop}
\end{eqnarray}
in terms of the parameter $\beta_R$ to be fixed below.

\subsection{Scalar meson contribution to the low--energy constants}

Meson--exchange in the $t$--channel can also contribute to the LECs as
pointed out in \cite{bmlet}. Denoting by $S$ and $S_1$ the scalar
octet and singlet with $M_S \simeq M_{S_1} \simeq 1$~GeV, respectively,
the lowest order effective Lagrangian coupling the scalars to the
Goldstone bosons and  to the baryon octet reads
\begin{eqnarray}
& & \!\!\!\!\!\!\!\!\!
{\cal L}_{\rm eff} (U,B,S)  = {\cal L}_0 (S) + {\cal L}_{\rm int} (U,S)
+ {\cal L}_{\rm int} (B,S)\, \, , \nonumber \\
& & \!\!\!\!\!\!\!\!\!
{\cal L}_0 (S) =  \frac{1}{2} {\rm Tr} \, ( \partial_\mu S \partial^\mu
S - M_S^2 S^2) + \frac{1}{2} \, {\rm Tr}( \partial_\mu S_1 \partial^\mu
S_1 - M_{S_1}^2 S_1^2) \, \, , \nonumber \\
& & \!\!\!\!\!\!\!\!\!
{\cal L}_{\rm int} (U,S) = c_d \, {\rm Tr} (S u_\mu u^\mu) + c_m \,
{\rm Tr} (S \chi_+) + \tilde{c}_d \, S_1 \,{\rm Tr} (u_\mu u^\mu) +
\tilde{c}_m \, S_1 \, {\rm Tr}(\chi_+) \, \, , \nonumber \\  
& & \!\!\!\!\!\!\!\!\!
{\cal L}_{\rm int} (B,S) = D_S \, {\rm Tr}(\bar B \lbrace S,B \rbrace )
+ F_S \, {\rm Tr}(\bar B [ S,B ] ) + D_{S_1} \,S_1 \, {\rm Tr}(\bar B B  )
\label{leffS}
\end{eqnarray}
where the coupling constants $D_S$, $F_S$ and $D_{S_1}$ have chiral 
dimension zero.
For the couplings of the scalars to the Goldstone bosons, we use the
notation of \cite{reso} and the parameters determined therein, i.e.
$|c_d| = 42$~MeV, $|c_m| = 32$~MeV with $c_d c_m >0$. The singlet
couplings can be related to these in the large--$N_c$ limit
 via $\tilde{c}_{d,m}
= \pm c_{d,m}/\sqrt{3}$ with $\tilde{c}_d \tilde{c}_m >0$. Similarly,
we have $D_{S_1} = \pm 2 D_s / \sqrt{3}$.     
In fact, it is very difficult to pin down the couplings $D_S, F_S $
and $D_{S_1}$.\footnote{In what follows, we neglect the singlet
fields. There effects can be absorbed in the pertinent coupling
constant of the octet fields.} We will therefore leave $F_S$ and $D_S$
as free parameters and determine them as described below. 

Evaluating now the diagrams shown in Fig.4, which are scattering graphs and
direct couplings of the scalar mesons to the operator $\chi_+$, and
using the lowest order effective Lagrangian Eq.(\ref{leffS}), one
finds only contributions to ${\cal L}^{(2)}_{\phi B}$. Contributions
to the LECs $d_i$ arise if one accounts for higher dimension operators
in ${\cal L}_{\rm int} (B,S)$. This would amount to a proliferation of
unknown coupling constants. We have refrained from including such
terms based on the observation that the scalar contributon to the
dimension two LECs is rather small \cite{bmlet}. This amounts to
\begin{eqnarray}
\label{lecsscal}
& & \!\!\!\!\!\!\!\!\!\!
    b_0^S  = -\frac{2}{3}C_m\, D_S + \tilde{C}_m \, D_{S_1} \, \, , \, \,  
    b_D^S  = C_m \, D_S \, \, , \, \,  
    b_F^S  = C_m \, F_S \, \, ,  \\  
& & \!\!\!\!\!\!\!\!\!\!
    b_1^S  = \frac{1}{2} C_d \, D_S \, \, , \, \,  
    b_2^S   = C_d \, F_S \, \, , \, \,  
    b_3^S  = \frac{1}{2} C_d \, D_S \, \, , \, \,  
    b_8^S  = -\frac{2}{3}C_d\, D_S + \tilde{C}_d \, D_{S_1} \, \, ,
\nonumber
\end{eqnarray}
with
\begin{equation}
C_{m,d} = \frac{c_{m,d}}{M_S^2} \, \, , \quad 
\tilde{C}_{m,d} = \frac{\tilde{c}_{m,d}}{M_{S_1}^2} \, \, .
\end{equation}
We have now assembled all pieces to discuss the numerical values of
the various LECs.

\subsection{Determination of the low--energy constants}

For pinning down the numerical values of the various LECs, we need
to know the values of the parameters $\beta_\Delta$, $\delta_\Delta$, 
$\beta_R$, $F_S$, $D_S$ and of the baryon octet mass in 
the chiral limit, $\krig m$. We will later show that it is mandatory
to keep all these parameters, i.e. using 
only  the decuplet and/or the scalar mesons to estimate the LECs does
not lead to a consistent description of the baryon masses and $\sigma$--terms.
We determine these parameters as follows. Within the framework of 
resonance exchange saturation of the LECs , the baryon masses take the form 
\begin{equation}
\!\!\!\!\!\!\!\!  m_B  =  \krig{m} + m_B^{(3)} 
+ \frac{1}{m_B} \, \lambda_B  
+ \beta \,  D_B^\beta
+ \delta \, D_B^\delta
+ \epsilon \, D_B^\epsilon + D_B^S 
\quad ,
\label{massreso}
\end{equation}
where we have lumped the ${\cal O}(m_q)$ and 
the ${\cal O}(m_q^2)$ corrections togther  
 (in the constants $D_B^\beta$ and $D_B^S$, respectively). 
We remark that whenever possible, we have substituted 
the octet mass in the chiral limit by the corresponding
physical mass since these differences are of higher order.
The $\lambda$ contributions are the
$1/ m$ insertions from ${\cal L}_{\phi B}^{(2)}$. Similarly,
the $\beta$ and $\epsilon$ terms stem from tadpole graphs with
insertions proportional to the low--energy constants $b_{0,D,F}$ and
$b_i$, respectively. The
$\delta$ terms subsume the contributions from  ${\cal L}_{\phi B}^{(4)}$,
these are proportional to the low--energy constants $d_i$. In these
last three terms, we have abbreviated
\begin{eqnarray}
\beta D_B^\beta &=& \beta_\Delta D_{B , \Delta}^\beta 
+ \beta_R D_{B,R}^\beta \, \, , \quad
\delta D_B^\delta = \delta_\Delta D_{B, \Delta}^\delta
+ \delta_R D_{B,R}^\delta \, \, , \quad \nonumber \\
\epsilon D_b^\epsilon &=& \epsilon_\Delta D_{B ,\Delta}^\epsilon
+ \epsilon_R D_{B, R}^\epsilon \quad . 
\end{eqnarray}
Finally, the terms of the type $D_B^S$ are the scalar meson
contributions to the mass (which could also be absorbed in the
coefficients $\beta$ and $\epsilon$, cf. Eq.(\ref{lecsscal})). 
The numerical values of the $\lambda_B$, $D^\beta_B, D^\delta_B$,
and $D^\epsilon_B$  are given in table~1,  
using $F = 0.5$, $D=0.75$, $D_R = 0.60$, $F_R = 0.11$ and
$F_\pi = 100 \,$MeV.
We see that the dominant terms at ${\cal O}(m_q^2)$ are indeed the
tadpole graphs with an insertion from ${\cal L}_{\phi B}^{(2, {\rm
br})}$ (this holds for the masses but not for $\sigma_{\pi N}(0)$).
These numbers are different form the ones given in Ref.\cite{bmlet}
due to the different definitions of the coefficients $\beta, \delta$
and $\epsilon$ and because of the Roper--octet contribution. 
The  $D_B^S$ are (in GeV) 
\begin{eqnarray}
D_N^S        & = & -0.013 \, D_S + 0.057 \, F_S \, \, , \quad  
D_\Lambda^S    =    -0.042 \, D_S - 0.019 \, F_S \, \, , \nonumber \\  
D_\Sigma^S   & = & +0.052 \, D_S + 0.019 \, F_S \, \, , \quad  
D_\Xi^S        =   -0.045 \, D_S - 0.076 \, F_S \, \, ,
\label{DBS}
\end{eqnarray}  
where we have  used the results of \cite{reso} for the scalar couplings and 
$M_S = 0.983$ GeV. The four equations for the octet masses 
\beq
m_B = m_B [\krig{m} , \beta_\Delta, \delta_\Delta, \beta_R, F_S, D_S ]
\,\,\, , \eeq
allow to
pin down four parameters, we choose $\delta_\Delta$, $\beta_\Delta$,
$F_S$ and $D_S$. A similar equation can be worked out for the
pion--nucleon $\sigma$--term,
\beqa
\sigma_{\pi N} (0) &=& \sigma_{\pi N}^{(3)} (0)
+\frac{1}{m_N}\lambda_\sigma 
+ \beta_\Delta \,D_\sigma^{\beta_\Delta} + \beta_R \,D_\sigma^{\beta_R}
+\delta_\Delta \,D_\sigma^{\delta_\Delta}+\delta_R \,D_\sigma^{\delta_R}
\nonumber \\
&& +
\epsilon_\Delta \,D_\sigma^{\epsilon_\Delta} + \epsilon_R 
\,D_\sigma^{\epsilon_R}
+ D_S \,D_\sigma^{D_S} + F_S \,D_\sigma^{F_S} \,\,\, , 
\label{sigfit}
\eeqa
or numerically with $\sigma_{\pi N} (0) =45\,$MeV 
\beq
0.077 = -0.244\, \beta_\Delta + 0.020 \, \delta_\Delta - 0.056
\,\beta_R + 0.001 \, D_S -0.002 F_S \,\, ,
\eeq
in appropriate units of GeV. We end up with 5 equations for six
unknown parameters. Choosing $\beta_R$ as the undetermined one, this
leads to
\beqa
\krig m       &=& +0.7673 + 0.3737 \, \beta_R \,\,\, , \,\,\,
\beta_\Delta   = -0.3321 - 0.1126 \, \beta_R \,\,\, ,\nonumber \\
\delta_\Delta &=& -0.1387 - 0.0232 \, \beta_R \,\,\, , \,\,\,
D_S            =  -5.8244 + 1.4484 \, \beta_R \,\,\, ,\nonumber \\
F_S           &=& -0.8974 - 12.715 \, \beta_R \,\,\, .
\label{paras}
\eeqa
Units are GeV for $\krig m$, GeV$^{-1}$ for $\beta_\Delta, \beta_R$ and 
GeV$^{-3}$ for $\delta_\Delta$ whereas $F_S$ and $D_S$ are 
dimensionless. The prefactors have the appropriate dimensions, for 
example $\krig m \,$[GeV] = $0.767\,$[GeV]
$+0.374\,$[GeV$^2$]$\beta_R\,$[GeV$^{-1}$].
We can now give a bound on $\beta_R$ based on the assumption
that the ratio of the decuplet contribution and of the Roper octet
one is of the same order for all LECs. Since the contributions to the
$b_i$ do not contain any unknown parameters, we can use their ratio
to estimate the
corresponding one from the Roper--octet and we thus conclude
\beq
\biggl|\frac{ D_{B,R}^{\beta} \, \beta_R}{ D_{B,\Delta}^{\beta} \, 
\beta_\Delta} \biggr| 
\sim \biggl|\frac{D_{B,R}^{\epsilon} \, \epsilon_R} 
{D_{B,\Delta}^{\epsilon} \, \epsilon_\Delta} \biggr|
\eeq
from which we derive the (soft) bound
\beq
|\beta_R | \le 0.1 \,\,\, {\rm GeV}^{-1} \quad. 
\label{rbound}
\eeq
Consequently, the scalar couplings $F_S$ and $D_S$ are falling into
the ranges $-2 < F_S < 0.5$ and $-6.0 < D_S < -5.7$. It would be interesting
to have some phenomenological bounds on these couplings as a check of
the consistency of our procedure.

Finally, we can now express the LECs in terms of the parameter
$\beta_R$,
\beqa
b_0 &=& -0.606 - 0.227 \, \beta_R \, , \,
b_D =   +0.079 + 0.014 \, \beta_R \, , \,
b_F =   -0.316 - 0.536 \, \beta_R \, , \nonumber \\
b_1 &=& -0.004 + 0.024 \, \beta_R \, , \,
b_2   = -0.187 - 0.421 \, \beta_R \, , \nonumber \\
b_3 &=& +0.018 + 0.024 \, \beta_R \, , \,
b_8   = -0.109 - 0.032 \, \beta_R \, , \nonumber \\
d_1 &=& +0.008 + 0.001 \, \beta_R \, , \,
d_2   = +0.035 + 0.006 \, \beta_R \, , \nonumber \\
d_3 &=& +0.069 + 0.012 \, \beta_R \, , \,
d_4   = -0.077 - 0.013 \, \beta_R \, , \nonumber \\
d_5 &=& -0.050 - 0.008 \, \beta_R \, , \,
d_7   = -0.018 - 0.003 \, \beta_R \, , \,
d_8   = -0.103 - 0.017 \, \beta_R \,  \, \, ,
\label{LECvals}
\eeqa
with the $b_i$ in GeV$^{-1}$ and the $d_i$ in GeV$^{-3}$ and the
coefficients have corresponding dimensions.
In table~2, we compare the LECs of the
dimension two meson--baryon effective Lagrangian with the values 
 previously determined from KN scattering 
data. We have transformed the results of Ref.\cite{norb} into our 
notation.   As can be seen from table~2, most (but not all)
coefficients agree in sign and magnitude.
We remark that the procedure used in \cite{norb} involves the summation of
arbitrary high orders via a Lippmann--Schwinger equation and is thus 
afflicted with some uncertainty not controled within
CHPT.\footnote{These values have recently been refitted taking account
also $\eta$ and kaon photoproduction data \cite{norb2}. 
The resulting values are
somewhat different from the ones given in ref.\cite{norb}.} For
comparison, we also give the values of $b_{0,D,F}$ from the $q^3$
calculation \cite{bkmz}.

We end this paragraph with some comments on estimating the LECs from
the decuplet or decuplet and Roper--octet alone. Consider first the
case of  the decuplet only. We then have three parameters determined 
by a least--square fit to the four masses. One finds $\krig m =
620\,$MeV, $\beta_\Delta = -0.551$~GeV$^{-1}$
and $\delta_\Delta = 0.975$~GeV$^{-3}$ and the
deviations of the fitted from the physical masses range from -154 to
121 MeV. This means that in particular the 
deviation from the Gell-Mann--Okubo relation, 
\begin{equation}
\Delta_{\rm GMO} = \frac{1}{4} \bigl(3m_\Lambda+m_\Sigma-2m_N-2m_\Xi
\bigr) \quad ,
\label{GMO}
\end{equation}
which empirically is about 6.5 MeV, is very large (of the order of
200 MeV). The LECs take values which are considerably larger than the
ones given above (cf. table~2 and Eq.(\ref{LECvals})). Furthermore,
the $\sigma$--term is completely fixed and turns out to be
$\sigma_{\pi N} (0) = 89\,$ MeV, considerably larger than the
empirical value. One could try to remedy the situation by adding the
Roper--octet, thus having one more free parameter at ones disposal. 
One can then solve the linear system of equations for  the four masses
in terms of the four parameters. This does, however, not lead to
sensible results. One finds $\beta_R \simeq 10$~~GeV$^{-1}$, 
much larger than the
bound given in Eq.(\ref{rbound}). This leads to absurd results like
$\krig m \simeq 3.9\,$GeV. It is therefore mandatory to include the
scalar meson exchange to consistently describe the scalar sector of
three flavor baryon CHPT.

\subsection{Two--loop contributions to the LECs}
\label{sec:twoloop}

As shown, the estimation of the LECs entering the baryon masses
involves Goldstone boson loop diagrams, compare Fig.~3. With such
graphs, one encounters a new mass scale which is non--vanishing in the
chiral limit, $\krig{m}_R - \krig{m} \ne \, 0$ as $M_\phi \to 0$. 
Therefore, a strict one--to--one correspondence between the 
expansion in small momenta and the number of Goldstone boson
loops is no longer guaranteed, as it is known from the study of
relativistic baryon CHPT \cite{gss}. A similar situation arises for
the calculation of the deviations from Dashen's theorem, which
involves photons coupled to heavy (axial--)vector mesons in the loops,
see e.g. Refs.\cite{devdash}. Again, the new mass scale (here the
vector meson mass in the chiral limit) spoils the strict power
counting. Indeed, it has been shown that two--loop graphs modify the
leading order results to the deviations form Dashen's 
theorem \cite{joachim}. We therefore also have to address this issue
in our context.

The possible two--loop graphs to be considered are shown in
Fig.~5. Here, we concentrate on the diagrams with the maximal number
of resonance propagators and the simplest $B B^* \phi$ coupling via
the chiral connection $\Gamma_\mu$. This means
for example in diagram (e) one of the intermediate propagators
is a nucleon and the other one refers to an excited intermediate
spin--3/2 or spin--1/2 state.  For all these graphs,
all possible combinations of decuplet, Roper--octet and nucleon 
propagators should be considered. From these, we only take the
leading ones in the large mass limit subject to the constraint
that the vertices involved are constructed from the chiral
connection.  The corresponding contributions from the axial--vector
couplings $\sim u_\mu$ involve other coupling constants and are
therefore algebraically independent from the ones considered here.
In fact, to simplify the algebra, we will focus 
on the case where all resonance propagators refer to the Roper--octet.
A more detailed account of these calculations is given in \cite{bora}.

Consider first the tadpole--type diagrams like in Fig.5a. Without 
Clebsch--Gordan coefficients and other prefactors, it takes the form
\beqa
I_{5a} &=& \int \frac{d^4 l}{(2\pi)^4} \int \frac{d^4 k}{(2\pi)^4} 
\frac{i}{l^2-M_a^2}\, (p\!\!/ - k \!\!/) \,\gamma_5 \frac{i}{k \!\!/ -
  m_R} \, (p\!\!/ - k \!\!/) \, \gamma_5 \, \frac{i}{(p-k)^2-M_b^2}
\nonumber \\
&=& -M_a^2 \biggl[ 2L + \frac{1}{16\pi^2} \ln \frac{M_a^2}{\lambda^2}
\biggr] \, \biggl\{ i m_R (m_R^2 + M_b^2) 
\biggl[ 2L + \frac{1}{16\pi^2} \ln 
\frac{m_R^2}{\lambda^2} \biggr] 
\nonumber \\
& & \qquad \qquad \qquad \qquad
+\frac{i}{32\pi^2}\frac{\krig m}{m_R^2} M_b^4 -
\frac{i}{16\pi^2m_R}M_b^4 \ln\frac{M_b^2}{\lambda^2} \biggr\} + \ldots
 \,\,\, , 
\label{Itadpole}
\eeqa
where we have only retained the leading terms in the expansion in the 
resonance mass $m_R$ as indicated by the ellipsis, i.e. we have
approximated $m_R - \krig{m} \simeq m_R$ and so on. We remark that the
disturbing term $\sim M_a^2 \ln M_a^2$ cancels in the sum of graphs
5a, its partner with the tadpole at the other vertex and graph 5f.
Taking into account the pertinent counter term graphs to renormalize the
infinities connected to the tadpole, we find
\beq
I_{5a} \sim M_a^2 \biggl\{ i m_R \,(m_R^2 + M_b^2) \,
 \biggl[ 2L + \frac{1}{16\pi^2} \ln
\frac{m_R^2}{\lambda^2} \biggr] + {\cal O}(M_b^4) \biggr\} \, \, .
\label{Itadpole2}
\eeq
The first term in Eq.(\ref{Itadpole2}) can be completely absorbed in
the renormalization constant $\alpha_R^{(2)}$ and the second one
leads to a new renormalization constant $\alpha_R^{(4)}$. Its finite piece
modifies the value of the finite dimension four 
one--loop contribution $\delta_R$.
Simarly, for graphs with intermediate decuplet states, to leading
order in the resonace mass on can absorb the two--loop contribution
entirely in a redefinition of the constants $\alpha_\Delta^{(2)}$ and
$\alpha_\Delta^{(4)}$. Furthermore, if one works out the first finite
contribution at orders $M_a^{2,4}$, one finds that these are numerically much
smaller than the corresponding pieces from the one loop diagram.
Similarly, we work out the
contribution from the graphs of the type 5d. The integral stripped off
all prefactors and expanded in powers of the resonance mass takes the
form
\beqa
I_{5d} &=&  
-2 m_R^3 \, \biggl[ 2L + \frac{1}{16\pi^2} 
\ln \frac{m_R^2}{\lambda^2} \biggr]^2 (M_a^2+M_b^2) \no \\ 
&& -2 m_R \,\frac{1}{16\pi^2} \ln\frac{m_R^2}{\lambda^2} 
 \biggl[ 2L + \frac{1}{16\pi^2} \ln\frac{m_R^2}{\lambda^2} \biggr]
\,(M_a^4+M_b^4) + \ldots
\label{Ihump}
\eeqa
Again, standard renormalization has to be performed 
and one is left with a contribution to $\alpha_\Delta^{(2)}$
and  one to $\alpha_\Delta^{(4)}$. The latter can be made to
vanish if one sets $\lambda= m_R$.

The other diagrams shown in Fig.5 can not be given in closed
analytical form. To get an estimate about their contributions, we
perform asymptotic expansions in the external momentum $p$ making use
of the formalism developed in Ref.\cite{daydy} (and references therein).
For external momenta (here $p = \krig{m}$) below the first threshold
(here $p_{\rm thr}= \krig{m} +2M_\pi$), one can expand around $p=0$ to
leading order. It is straightforward to show \cite{bora} that
this procedure is sufficient to estimate the leading order terms in
the large resonance mass expansion (here $m_R$). Specifically, we have
\beq
I_{5\alpha}(p^2) = I_{5\alpha}(0) + \frac{p^2}{2d}\, \Box_p 
\,I_{5\alpha}(p^2) + {\cal O}(p^4) \quad (\alpha = b,c,e) \,\,\, ,
\label{asyexp}
\eeq
in $d$ space--time dimensions. The explicit formulae for the various
contributions are very lengthy and can be found in \cite{bora}. Here,
we just show a typical result after picking out the leading terms in
the $m_R$ expansion,
\beq
I_{5c} = e_1 \, m_R^5 + e_2 \, m_R^3 \, ( M_a^2 + M_b^2) + e_3 \, m_R 
( M_a^4 + M_b^4) + e_4 \, m_R M_a^2 M_b^2 + \ldots 
\,\, \, \, ,
\eeq
with $\epsilon = d-4$ and the coefficients $e_i$
contain divergences. After removing the divergent pieces
$\sim 1/\epsilon$ and $1/\epsilon^2$, we are essentially left with 
contributions to the renormalization constants 
$\alpha_\Delta^{(2,4)}$ and $\alpha_R^{(2,4)}$, i.e. the two--loop
effects can be completely absorbed in pertinent redefinitions.
Only the finite one--loop constant $\delta_R$ is modified. The corresponding 
tree graph contributions to the baryon masses and $\sigma$--terms 
at order $M_\phi^4$ from ${\cal L}_{\phi B}^{(4)}$ are, however,
very small and thus an accurate knowledge of this coupling is not
needed. From this we conjecture that a similar mechanism is also
operative at higher orders and that our one--loop approach to estimate
the baryon resonance excitations to the various LECs is a consistent
procedure.

\subsection{Some two--loop contributions to the baryon masses}
\label{sec:masstwo}

Having worked out the two--loop contributions to the LECs $b_i$ and
$d_i$ in paragraph~\ref{sec:twoloop}, we can use the same machinery to
estimate some typical two--loop, i.e. order $q^5$, contributions to
the baryon masses. This should only considered indicative since we do
not attempt a full ${\cal O}(q^5)$ calculation here, which besides all
two--loop diagrams would also involve one--loop graphs with exactly
one insertion from ${\cal L}_{\phi B}^{(3)}$.

Consider first the tadpole--type graph like in Fig.5a
(and its partner with the tadpole on the other side) but with the
essential difference that the intermediate propagator refers to a 
groundstate spin--1/2 state in the heavy baryon formalism. Using the
appropriate Feynman rules, the momentum--space integral $I_t$ is
\beq
I_t = \int \frac{d^4 q}{(2\pi)^4} \int \frac{d^4 k}{(2\pi)^4} 
\frac{i}{q^2-M_a^2}\, \frac{i}{k^2-M_b^2}\,\frac{i}{v \cdot (p-k)}
(S \cdot k)^2 \,\,\, .
\label{It}
\eeq
To leading order, we can neglect the baryon off--shell momentum. $I_t$
then takes the simple form
\beq
I_t = \frac{1}{4} \ f(M_a) \, M_b^2 \, \int \frac{d^4 k}{(2\pi)^4} 
\frac{-1}{[k^2-M_b^2] \, v \cdot k} 
= \frac{-i}{32\pi} \, M_a^2 \, M_b^3 \biggl[2L + 
\frac{1}{16\pi^2} \ln \frac{M_a^2}{\lambda^2} \biggr] \,\,\, .
\label{Itres}
\eeq
As expected from the power counting, this contributions starts at
order $M_\phi^5$. Renormalizing the divergence  
and restoring the appropriate prefactors, we have
the following contribution to the nucleon mass from these tadpole--type
graphs
\beq
m_N^{(5,a)} = \frac{45F^2+5D^2-6DF}{3072\, \pi^3\,F_\pi^4} 
\,M_K^5 \, \ln\frac{M_K^2}{\lambda^2} \,\,\, ,
\label{mNtad}
\eeq
where we have set for simplification $M_\pi=0$ and $M_K = M_\eta$.
Similarly, we can work out the contribution from the analog of fig.~5f 
with heavy baryon propagators,
\beq
m_N^{(5,f)} \simeq -\frac{45F^2+17D^2 -30DF }{768\, \pi^3 \,F_\pi^4} \,
M_K^5 \, \ln\frac{M_K^2}{\lambda^2} \,\,\, .
\label{mNbutt}
\eeq
The ``double--hump'' graph (compare fig.~5d) vanishes to this order,
$m_N^{(5,d)} = 0$. Eqs.(\ref{mNtad},\ref{mNbutt}) 
will be used in the next section to get an
order of magnitude estimate of the two--loop corrections to the
nucleon mass.


\section{Results and discussion}
\label{sec:results}

\subsection{Results for the central values of resonance parameters}

In this section, we discuss the results for our central values of
parameters, setting $\beta_R=0$ (varying $\beta_R$ within its allowed range,
$-0.1 \le \beta_R \le 0.1$ GeV$^{-1}$, 
does only lead to irrelevant changes).  These are
$F_\pi =100 \, {\rm MeV} \,   ,  \, D = 0.75 \,  ,  \, F = 0.5 \,   ,  \, 
{\cal C} = 1.5 \,  , \, D_R = 0.6  \,  , \, F_R = 0.11$
and we set $\lambda =1 \,$GeV. The various mesonic LECs $L_i^r
(\lambda)$ are the central values taken from the compilation of
Bijnens et al. in Ref.\cite{daphne}.  For these central values, 
all baryon masses are fitted exactly, together with $\sigma_{\pi N} (0) 
=45\,$MeV. The theoretical uncertainties
induced by the spread of these parameters, in particular due
to the $\pm 10\,$MeV uncertainty in  $\sigma_{\pi N} (0)$ \cite{gls},
will be discussed in the next subsection.
We find for the octet baryon mass in the
chiral limit using Eq.(\ref{paras})
\begin{equation}
\krig{m} = 767 \, \, {\rm MeV} \quad . 
\label{mkrig}
\end{equation}
The quark mass expansion of the baryon masses, in the notation of 
Eq.(\ref{massform}), reads
\begin{eqnarray}
m_N & = & \krig{m}  \, ( 1 + 0.34 - 0.35 + 0.24 \, ) \, \, ,
\nonumber \\
m_\Lambda & = & \krig{m} \, ( 1 + 0.69 - 0.77 + 0.54 \, ) \, \, ,
\nonumber \\
m_\Sigma & = & \krig{m} \, ( 1 + 0.81 - 0.70 + 0.44 \, ) \, \, ,
\nonumber \\
m_\Xi & = & \krig{m} \, ( 1 + 1.10 - 1.16 + 0.78 \, ) \, \, .
\label{mexpand}
\end{eqnarray}
We observe that there are large cancellations between the second order
and the leading non--analytic terms of order $q^3$, a well--known effect.
The fourth order
contribution to the nucleon mass is fairly small, whereas it 
is sizeable for the $\Lambda$, the $\Sigma$ and the $\Xi$. This is
partly due to the small value of $\krig{m}$, e.g. for the $\Xi$ the
leading term in the quark mass expansion gives only about 55\% of the
physical mass and the second and third order terms cancel almost completely.
The fourth order
contributions are indeed dominated by the one--loop graphs with
insertions from ${\cal L}^{(2, \rm br)}_{\phi B}$ as conjectured by
Jenkins and Manohar \cite{liz,dobo}. However, one can not 
neglect the terms with insertions from the remaining dimension two
terms, which are proportional to the $b_i$ and stem from relativistic
$1/m$ corrections. In contrast, the contributions
from the local terms ${\cal L}^{(4)}_{\phi B}$ are fairly small,
i.e. one does not need to know the LECs $d_i$ very accurately. 
From the chiral expansions exhibited in Eq.(\ref{mexpand}) one can not
yet draw a final conclusion about the rate of covergence in the
three--flavor sector of baryon CHPT. Certainly, the breakdown of CHPT
claimed in ref.\cite{juerg} is not observed. On the other hand, the
conjecture \cite{jms}  that only the leading non--analytic corrections (LNAC)
$\sim m_q^{3/2}$ are large and that further terms like the ones $\sim m_q^2$
are moderately small, of the order of 100 MeV, is not supported by our
findings. 

We now turn to the $\sigma$--terms and the strangeness content of the
nucleon. The pion--nucleon $\sigma$--term is used in the fitting
procedure. It is, however, 
instructive to disentangle the various contributions to $\sigma_{\pi
  N} (0)$ of order $q^2$, $q^3$ and $q^4$, respectively,
\begin{equation}
\sigma_{\pi N} (0) = 58.3 \, ( 1 - 0.56  + 0.33) \, \, \, {\rm MeV} 
= 45 \, \, {\rm MeV}     \, \, ,
\label{signo}
\end{equation}
which shows a moderate convergence, i.e. the terms of increasing order
become successively smaller. Still, the $q^4$ contribution is
important. Also, at this order no $\pi \pi$ rescattering effects are
included. Rewriting the $\sigma$--term as \cite{juerg}
\beq 
\sigma_{\pi N} (0) = \frac{\hat \sigma}{1-y}
\label{sighat}
\eeq
we find for the strangeness fraction $y$ and for $\hat \sigma$
\beq
y = 0.21 \,\,\,, \quad \hat \sigma = 36 \, {\rm MeV} \,\,\, .
\eeq
The value for $y$ is within the band deduced in ref.\cite{gls}, $y =
0.15 \pm 0.10$ and the value for $\hat \sigma$ compares favourably
with Gasser's estimate, $\hat \sigma = 33 \pm 5\,$MeV \cite{juerg}.

Finally, we consider the kaon--nucleon $\sigma$--terms and the various
scalar form factors. As stressed in section~\ref{sec:sigmaterms},
there appear undetermined renormalization constants at order $q^4$
as long as one works in the isospin limit $m_u=m_d$. These are
expected to be of order one. Indeed, for the pion--nucleon
$\sigma$--term one can calculate this constant (called $a_\pi$) since the full
renormalization has been performed ($\beta_R = 0$), 
\beq
\sigma_{\pi N} (0) = (50.1 - 14.4 \, a_\pi ) \, {\rm MeV} \rightarrow
a_\pi = 0.36 \quad ,
\eeq
i.e. $a_\pi$ has indeed the expected size. For the kaon--nucleon
$\sigma$--terms, we can give the results up to two constants (called
$a_{K1}$ and $a_{K2}$),
\beqa
\sigma_{KN}^{(1)} (0) &=& (369 - 306 \, a_{K1}) \,\, {\rm MeV} \,\,
, \nonumber \\
\sigma_{KN}^{(2)} (0) &=& (934 - 437 \, a_{K2}) \,\, {\rm MeV}\,\, ,
\eeqa
with the respective chiral expansions
\beqa
\sigma_{KN}^{(1)} (0) &=& (528 - 524 + 365 
- 306 \, a_{K1}) \,\, {\rm MeV} \,\, , \nonumber \\
\sigma_{KN}^{(2)} (0) &=& (290 - 49 + 693 
-437 \, a_{K2}) \,\, {\rm MeV}\,\, ,
\label{KNsig}
\eeqa
where the terms refer to the orders $q^2$, $q^3$, $q^4$ and $q^4$, 
respectively (the $q^4$ contribution independent of the
renormalization constant $a_{Ki}$ is shown separately). These numbers
agree with the $q^3$ calculation of ref.\cite{bkmz} (to that order). 
Varying $a_{K1}$ and  $a_{K2}$ between 0.5 and 1, we have
\beq
\sigma_{KN}^{(1)} (0) = 73 \ldots 216 \,\, {\rm MeV} \,\, , \quad
\sigma_{KN}^{(2)} (0) = 493 \ldots 703 \,\, {\rm MeV}\,\, .
\eeq
These numbers should be considered indicative and can only be
sharpened in a  calculation with $m_u \neq m_d$. For a discussion of
the various $\sigma$--term shifts to the pertinent Cheng--Dashen
points, we refer to ref.\cite{bora}.


\subsection{Theoretical uncertainties}

In the previous paragraph, we gave the results for the central
values of the input parameters. Here, we will discuss the spread of
the results due to the uncertainties related to these numbers. 

Consider first the dependence on the coupling constant ${\cal C}$ and
the values of the axial--vector couplings $F$ and $D$. For comparison
with our central values, we also use $|{\cal C}| =1.2$ determined by
Butler et al. \cite{bss}, $|{\cal C}| =1.8$ from the decay $\Delta \to
N\gamma$ and $D= 0.85 \pm 0.06$, $F= 0.52 \pm  0.04$
given by Luty and White \cite{luwh}. The results depend only very
weakly on these parameters, i.e. they vary within a few percent. For
the case of ${\cal C}$, this weak dependence stems from the fact that
${\cal C}$ only changes the value of $\epsilon_\Delta$ whereas the
changes in the much more important $\beta_\Delta$ and $\delta_\Delta$
are absorbed in the new fit values of
$\alpha_\Delta^{(2,4)}$. The weak dependence on the actual values of
$F$ and $D$ is due to compensating contributions of third and fourth 
order and the already mentioned dominance of the tadpole graphs
with the symmetry breakers $\sim b_{0,D,F}$.
Consider now variations in the renormalization scale $\lambda$. 
The latter dependence is induced since we estimate the LECs 
from resonance exchange and would
disappear once all LECs could be determined from data. In table~3
we show results for the range $0.8 \, {\rm GeV} \le \lambda \le 1.2\,$
GeV,  for the central values of $F_\pi$,
$F,D,F_R$ and $D_R$. The strangeness fraction $y$ is most
notably affected. The chiral series for the masses converges quicker
for lower values of $\lambda$. In table~4, we vary the value of 
$\sigma_{\pi N}(0)$ generously by $\pm 10\,$MeV. Again, the
strangeness fraction shows the largest variation. All these variations
are essential linear in $\delta\sigma_{\pi N}(0)$.
Finally, we remark that varying the pseudeoscalar decay constant
$F_\pi$ between 93 and 113 MeV does also not alter any of the previous
numbers drastically.  We therefore assign the following theoretical
uncertainty to the results for the average octet mass in the chiral
limit, the strangeness fraction  $y$ and $\hat \sigma$, in order
\begin{equation}
\krig m = 767 \pm 110 \, \, {\rm MeV} \, \, , \, \,
y = 0.21 \pm 0.20 \, \, , \, \,
\hat \sigma = 36 \pm 7 \, \, {\rm MeV} \, \, .
\label{valunc}
\end{equation}
These uncertainties do not include the possible effects of higher
orders which can only be assesed if one performs such a calculation.
As an indication of  genuine two loop contributions, we quote the
results of the diagrams evaluated in paragraph~\ref{sec:masstwo}.
We find (setting $\lambda = 1\,$GeV) 
\beq 
m_N^{(5a)} = -52 \,\,{\rm MeV} \,\, \quad
m_N^{(5f)} = -13 \,\,{\rm MeV} \,\, ,
\eeq
which are individually very small, i.e. well within the uncertainties
discussed above. It is expected that the major contribution at
order $q^5$ does indeed come from one--loop graphs with insertions
from ${\cal L}_{\pi N}^{(3)}$ and not from the genuine two loop
diagrams. To quantify this statement, such an order $q^5$ calculation
has to be performed. That, however, goes beyond the scope of the
present paper.


\section{Summary and outlook}
\label{sec:summary}

In this paper, we have considered the chiral expansion of the 
groundstate baryon octet baryon masses and the pion--nucleon
$\sigma$--term to quadratic (fourth) order in the quark (Goldstone boson)
masses, in the framework of heavy baryon chiral perturbation
theory. The pertinent results of this investiagtion can be summarized
as follows:
\begin{enumerate}
\item[$\circ$]We have constructed the most general effective
Lagrangian to fourth order in the small parameter $q$ (external
momentum or meson mass) necessary to investigate the scalar sector. Besides the
standard dimension two symmetry--breaking terms, Eq.(\ref{leff2st}), it
contains further dimension two operators with derivatives acting on
the Goldstone boson fields and kinematical $1/m$ corrections,
cf. Eq.(\ref{leff2der}) and Eq.(\ref{relcorr}).  
\item[$\circ$]We have given the complete expressions for the baryon
masses ($m_{N,\Lambda,\Sigma,\Xi}$) at order $q^4$ 
togther with the pion--nucleon $\sigma$--term.
At this order, divergences appear (in contrast to ${\cal O}(q^3)$,
where the one loop corrections to the masses and $\sigma_{\pi N}(0)$
are finite). The renormalization procedure to render the baryon
self--energies and also $\sigma_{\pi N}(0)$ finite (by use of the
Feynman--Hellmanm theorem) involves additional terms as listed in 
Eq.(\ref{leffrenor}). The renormalization of the kaon--nucleon
$\sigma$--terms and the corresponding scalar form factors is further 
complicated by additional momentum--(in)dependent divergences, as
detailed in appendix~\ref{app:renorm}.
\item[$\circ$]There appear seven low--energy constants (LECs) at order $q^2$
(called $b_i$) and  seven more at order $q^4$ (called $d_i$) to
calculate the masses and pion--nucleon $\sigma$--term. 
Two of the latter amount to quark mass renormalizations of two of the
$b_i$. Since there do not exist enough data to fix all these, we have
estimated them by means of resonance exchange. Besides standard tree
graphs with scalar meson exchange, this involves Goldstone
boson loops with intermediate baryon resonances (spin--3/2 decuplet
and the spin--1/2 (Roper) octet)
for the scalar--isoscalar LECs. We have discussed a
consistent scheme how to implement resonance exchange under such
circumstances, i.e. which avoids double--counting and abids to the
strictures from analyticity. Within the one--loop approximation and to
leading order in the resonance masses, the analytic pieces of the
pertinent graphs are still divergent, i.e.
one is left with three a priori undetermined renormalization constants
($\beta_\Delta$, $\delta_\Delta$ and $\beta_R$). These have to be
determined togther with the finite scalar couplings $F_S$ and $D_S$
and the octet mass in the chiral limit. Using the baryon masses and
the value of $\sigma_{\pi N} (0)$ as input, we can determine all LECs
in terms of one parameter, $\beta_R$. We derive a bound on this parameter
and show that observables are insensitive to variotions of $\beta_R$
within its allowed range. We have also demonstrated
that the effects of two (and higher) loop diagrams can almost entirely
be absorbed in a redinition of the one loop renormalization
parameters.
\item[$\circ$]Within this scheme of estimating the LECs we determine
the baryon mass in the chiral limit, denoted $\krig
m$, $\krig{m} = 770 \pm 110\,$ MeV (accounting also for the uncertainty in
certain input parameters like e.g. the pion--nucleon $\sigma$--term). 
For the strangeness fraction $y$ we find
$y = 0.21 \pm 0.20$, consistent with dispersion--theoretical 
determinations. This translates into $\hat \sigma = 36 \pm 7\,$MeV,
compare Eq.(\ref{sighat}), in good agreement with a previous
calculation \cite{juerg}.
\item[$\circ$]The chiral expansions for the nucleon mass and the
pion--nucleon $\sigma$--term are moderately well behaved whereas
for the hyperon masses $m_{\Lambda,\Sigma,\Xi}$ there still appear
sizeable corrections at order $q^4$. This is partly due to the almost
complete cancelations between the terms of order $m_q$ and $m_q^{3/2}$
and the smallness of the baryon mass in the chiral limit. A definite
statement about the convergence of three--flavor baryon CHPT can thus
not yet be made. 
\item[$\circ$]We have also estimated the two kaon--nucleon
$\sigma$--terms, which take the ranges given in Eq.(\ref{KNsig})
based on dimensional analysis for the appearing renormalization 
constants $a_{K1,K2}$.

\end{enumerate}


\section*{Acknowledgements}
We are grateful to J\"urg Gasser and Joachim Kambor for some 
very useful comments. A series of discussions with various
participants of the workshop ``The Standard Model at Low Energies''
at the ECT* in Trento in May 1996 helped to sharpen our views. 
We are grateful to the ECT* for offering this opportunity.


\appendix
\def\theequation{\Alph{section}.\arabic{equation}}
\setcounter{equation}{0}
\section{Baryon mass coefficients at order $q^4$}
\label{app:mbcoeffs}

Here, we give the coefficients $\epsilon_{i,B}^{P(Q)}$ with
$i=1,2,3,4$ and $P,Q = \pi, K ,\eta$ for the $\Lambda$, the $\Sigma$
and the $\Xi$. All $d_i$ are understood as $d_i^r (\lambda)$.

\noindent $\Lambda:$
\begin{eqnarray}
\epsilon_{1,\Lambda}^\pi &=& -4(4d_3+\frac{8}{3}d_4+d_7+3d_8) \, \, , 
\epsilon_{1,\Lambda}^K =  -16(\frac{8}{3}d_3+\frac{2}{3}d_4
                    +d_7+d_8) \, \, , \nonumber \\
\epsilon_{2,\Lambda}^{\pi K} &=& 16(-d_7+ d_8+\frac{8}{3}d_3+\frac{4}{3}d_4) 
\, \, , \, \, 
\epsilon_{3,\Lambda}^\pi = 2b_D + 6b_0 - 4b_3-6b_8 
- \frac{D^2}{\krig{m}} \, \, , \nonumber \\
\epsilon_{3,\Lambda}^K & = & \frac{20}{3}b_D  + 8b_0 -12b_1 
-\frac{4}{3}b_3-8b_8 
- \frac{D^2  +9F^2}{3 \krig{m}} \, \, , \nonumber \\
\epsilon_{3,\Lambda}^\eta & = & 2b_0 - 4b_3 -2 b_8 
- \frac{D^2}{3 \krig{m}} \, \, , \, \,
\epsilon_{4,\Lambda}^{\pi \pi} = -16 \, D^2 \, b_D 
\,\, \, , \, \epsilon_{4,\Lambda}^{\pi K} = 16\, D^2 \, b_D
\,\, \, , \nonumber \\
\epsilon_{4,\Lambda}^{K \pi} & = & -(D^2 +9F^2) \, (\frac{4}{3}b_d
-2b_0) - (D-3F)^2 \, (-2 b_F+ b_0) \nonumber \\ 
& & \qquad \qquad \qquad\qquad \qquad 
- (D+3F)^2 (2 b_F + b_0)
\,\, \, , \nonumber \\
\epsilon_{4,\Lambda}^{K K} & = & (D^2+9F^2) \, \biggl(\frac{16}{3}b_D
 + 4b_0 \biggr) -
2 (D-3F)^2 \, (b_0+b_D+b_F) \nonumber \\ 
& & \qquad \qquad \qquad \qquad \qquad - 2(D+3F)^2 (b_0 +b_D-b_F)
\,\, \, , \nonumber \\
\epsilon_{4,\Lambda}^{\eta \pi} & = & -\frac{14}{9}b_D 
\,\, \, , \, \, 
\epsilon_{4,\Lambda}^{\eta K}  =  \frac{32}{9}b_D  \, \, \, \, .
\label{epslambda}
\end{eqnarray} 

\noindent $\Sigma:$
\begin{eqnarray}
\epsilon_{1,\Sigma}^\pi &=& -4(4d_3++d_7+3d_8) \, \, , 
\epsilon_{2,\Sigma}^{\pi K} = 16(-d_7+d_8) \, \, , 
\epsilon_{1,\Sigma}^K =  -16(d_7+d_8) \, \, , \nonumber \\
\epsilon_{3,\Sigma}^\pi &=& 6b_D  + 6b_0 - 8b_1  - 4b_3-6b_8 
- \frac{D^2+6F^2}{3 \krig{m}} \, \, , \nonumber \\
\epsilon_{3,\Sigma}^K & = & 4b_D  + 8b_0 - 4b_1 - 4b_3 - 8b_8 
- \frac{5D^2 + 6DF + 9F^2}{6 \krig{m}} \, \, , \nonumber \\
\epsilon_{3,\Sigma}^\eta & = & 2b_0 - \frac{4}{3}b_3 -2b_8 
- \frac{D^2}{3\krig{m}} \, \, , 
\epsilon_{4,\Sigma}^{\pi K} = \frac{16}{3}\, D^2\, b_D \,\, ,
\nonumber \\
\epsilon_{4,\Sigma}^{\pi \pi} & = &  (\frac{2}{3}D^2+4F^2) \, (6b_D+3b_0)
-4D^2 \, (-\frac{1}{3}b_D + \frac{1}{2}b_0) -12(2b_D+b_0) \,\, ,
\nonumber \\
\epsilon_{4,\Sigma}^{K \pi} & = & 6(D^2 + F^2) \, (2b_D+b_0) -
3 (D+F)^2 \, (-2b_F+b_0) 
- 3(D-F)^2 (b_0 +2b_f) \,\, \, , \nonumber \\
\epsilon_{4,\Sigma}^{K K} & = & 12(D^2+F^2) \, b_0 - 6(D+F)^2 \, (b_0+b_D+b_F) 
- 6(D-F)^2 (b_0 +b_D-b_F) \,\, \, , \nonumber \\
\epsilon_{4,\Sigma}^{\eta \pi} & = & \frac{2}{3} \,b_D 
\,\, \, , 
\epsilon_{4,\Sigma}^{\eta K} = 0 \, \, \, \, .
\label{epssigma}
\end{eqnarray} 

\noindent $\Xi:$
\begin{eqnarray}
\epsilon_{1,\Xi}^\pi &=& -4(4d_1-2d_5+d_7+3d_8) \, \, , 
\epsilon_{1,\Xi}^K = -16(d_1+d_2+d_3+d_5+d_7+d_8) \, \, , \nonumber \\
\epsilon_{2,\Xi}^{\pi K} & = & 8(4d_1+2d_2+d_5-2d_7+2d_8) \, \, , \nonumber \\
\epsilon_{3,\Xi}^\pi &=& 3(b_D - b_F + 2b_0 - b_1 +b_2 -b_3- 2 b_8) 
- \frac{3\,(D-F)^2}{4 \krig{m}} \, \, , \nonumber \\
\epsilon_{3,\Xi}^K & = & 2(3b_D + b_F + 4b_0 - 3b_1 -b_2 -3b_3-4b_8) 
- \frac{5D^2 + 6DF + 9F^2}{6 \krig{m}} \, \, , \nonumber \\
\epsilon_{3,\Xi}^\eta & = & 2b_0 - 3b_1 - b_2 -\frac{1}{3}b_3-2b_8 
- \frac{(D+3F)^2}{12 \krig{m}} \, \, , 
\epsilon_{4,\Xi}^{\pi \pi} = 0 \,\, \, , 
\epsilon_{4,\Xi}^{\pi K} = 0 \,\, \, , \nonumber \\ 
\epsilon_{4,\Xi}^{K \pi} & = & (5D^2 + 6DF +9F^2) \, (-2b_F+b_0) -
\frac{9}{2} (D+F)^2 \, (2b_D+b_0)
\nonumber \\ 
& & \qquad \qquad \qquad\qquad \qquad 
- \frac{1}{6}(D-3F)^2 (3b_0 -2b_D) \,\, \, , \nonumber \\
\epsilon_{4,\Xi}^{K K} & = & 2(5D^2+6DF+9F^2) \, (b_D+b_F+b_0) -
9 (D+F)^2 \, b_0  \nonumber \\ 
& & \qquad \qquad \qquad\qquad \qquad 
- \frac{1}{3}(D-3F)^2 (3b_0 +4b_D)
\,\, \, , \nonumber \\
\epsilon_{4,\Xi}^{\eta \pi} & = & -b_D - \frac{5}{3} \,b_F
\,\, \, , \epsilon_{4,\Xi}^{\eta K}  =  \frac{8}{3} \, (b_D +b_F)
\, \, \, \, .
\label{epsxi}
\end{eqnarray} 


\setcounter{equation}{0}
\section{Renormalization of the kaon--nucleon $\sigma$--terms 
and scalar form factors}
\label{app:renorm}

Here, we discuss the renormalization related to the two kaon--nucleon
$\sigma$--terms and the three scalar form factors $\sigma_{\pi N} (t)$
and $\sigma_{KN}^{(1,2)} (t)$. Since we are dealing with composite 
operators, we use the standard procedure and generalize the
original Lagrangian used to calculate the baryon masses,
${\cal L} = \sum_i {\cal L}_i$, in the following way
\beq
{\cal L} \to  {\cal L}+ {\sum_{q,i}}' \, c_{q,i} \, \biggl( 
-\frac{\partial {\cal L}_i}{\partial m_q} \biggr)
\,\, , \quad q=u,d,s \,\, ,
\eeq
where the $c_{q,i}$ are sources and the prime on the sum indicates that
not all values of $q$ and $i$ are necessarily taken.
Consider now a diagram labelled 'j',
which leads to  a divergent contribution to the $\sigma$--term. The
renormalization for that graph proceeds via
\beq
-\frac{\partial {\cal L}_i}{\partial m_q} \to
-\frac{\partial {\cal L}_i}{\partial m_q} + \frac{1}{F_\pi^2} \, L \,
{\sum_{k,q'}}' \, f_{q',k,j} \, \biggl(-\frac{\partial {\cal L}_k}
{\partial m_{q'}} \biggr)\,\, ,
\eeq
with the $f_{q',k,j}$ are Clebsch--Gordan coefficients.
To order $q^4$, there are two types of divergences, namely $\sim
L\,M^2\,t$ and $\sim L \, M^4$. The $t$--dependent ones can be renormalized
with the help of two counter terms of the form
\beq
{\cal L}_1 = {\rm Tr}(\bar B \{\chi_+, [D_\mu,[D^\mu,B]]\}) \,\, ,\,\,
{\cal L}_2 = {\rm Tr}(\bar B [ \chi_+, [D_\mu,[D^\mu,B]] ]) \,\, .
\eeq
For the $t$--independent ones, one needs six independent terms,
\beqa
{\cal L}_3 &=& {\rm Tr}(\bar B [ \chi_+,[\chi_+, B] ] ) \,\, ,\,\,
{\cal L}_4 = {\rm Tr}(\bar B \{\chi_+,[\chi_+, B]\} ) \,\, , \, \,
{\cal L}_5 = {\rm Tr}(\bar B \{\chi_+,\{\chi_+, B\}\} ) \,\,
\nonumber \\
{\cal L}_6 &=& {\rm Tr}(\bar B \chi_+) \, {\rm Tr}( \chi_+ B) \,\, ,\,\,
{\cal L}_7 = {\rm Tr}(\bar B [\chi_+,B]) \, {\rm Tr}( \chi_+) \,\, ,\,\,
{\cal L}_8 = {\rm Tr}(\bar B B) \, {\rm Tr}( \chi_+)\,
{\rm Tr}( \chi_+) \,\, .\nonumber \\
&& \,\,\,\,
\eeqa
Consider as an example the composite operator $\bar u u$ and the
corresponding matrix element $<p'|\bar u u|p>$. A specific contribution
to it is given  by the diagram in fig.~2
with an insertion of the $b_1$--vertex, leading to (cf. Eq.(\ref{Isigt})),
and neglecting for simplicity the $\pi^0-\eta$ mixing,
\beq
I_\sigma (t) = i \, (4M_a^2 -t ) \, L + {\rm finite} \,\,\,{\rm terms} 
\eeq
Expanding the operator $\bar u u$ and the $b_1$--term of the effective
Lagrangian to second order in the meson fields, the divergent parts from
$I_\sigma$ contribute as follows
\beq
<p'|\bar u u|p>_{\rm div} = -\frac{8\,b_1}{F_\pi^2}\,B\,L\,
\biggl(\frac{3}{2}M_\pi^2 + 2M_K^2 + \frac{1}{2}M_\eta^2 \biggr) 
+ \frac{8\,b_1}{F_\pi^2}\,B\,L\,t \,\, .
\eeq
The $t$--dependent divergence is renormalized with the help of
${\cal L}_1$ via
\beq
-\frac{2\,b_1}{F_\pi^2}\,\,L\, \biggl\{ 
-3 \frac{\partial {\cal L}_1}{\partial m_u}
-2 \frac{\partial {\cal L}_1}{\partial m_d}
- \frac{\partial {\cal L}_1}{\partial m_s} \biggr\}
\eeq
i.e . for the Clebsch--Gordan coefficients
\beq
f_{u,1,j} = -6\, b_1 \,\, , \, \,
f_{d,1,j} = -4\, b_1 \,\, , \, \,
f_{s,1,j} = -6\, b_1 \,\, , 
\eeq
and zero otherwise. The $t$--independent divergences in the
matrix element $<p'|\bar u u|p>$ are renormalized with the help of
${\cal L}_3$, ${\cal L}_5$, ${\cal L}_6$ and ${\cal L}_8$,
\beq
-\frac{b_1}{2\,F_\pi^2}\,\,L\, \biggl\{ 
-\frac{5}{12} \frac{\partial {\cal L}_3}{\partial m_u}
-\frac{5}{12} \frac{\partial {\cal L}_3}{\partial m_d}
-\frac{9}{8} \frac{\partial {\cal L}_5}{\partial m_u}
-\frac{5}{8} \frac{\partial {\cal L}_5}{\partial m_d}
-\frac{1}{8} \frac{\partial {\cal L}_5}{\partial m_s}
+2 \frac{\partial {\cal L}_6}{\partial m_u}
-\frac{1}{2} \frac{\partial {\cal L}_8}{\partial m_u} \biggr\}
\eeq
with  the Clebsch--Gordan coefficients
\beqa
f_{u,3,j} &=& -\frac{5}{24}\, b_1 \,\, , \, \,
f_{d,3,j} = -\frac{5}{24}\, b_1 \,\, , \, \,
f_{u,5,j} = -\frac{9}{16}\, b_1 \,\, , \, \,
f_{d,5,j} = -\frac{5}{16}\, b_1 \,\, , \nonumber \\
f_{s,5,j} &=& -\frac{1}{16}\, b_1 \,\, , \, \,
f_{u,6,j} =   b_1 \,\, , \, \,
f_{u,8,j} = -\frac{1}{4}\, b_1 \,\, .
\eeqa


\setcounter{equation}{0}
\section{$\sigma_{\pi N}(0)$ and strangeness fraction coefficients at
  ${\cal O}(q^4)$}
\label{app:sigmaeps}

Here, we collect the coefficients $\epsilon_{i,\sigma}^{P(Q)}$ and
$\epsilon_{i,y}^{P(Q)}$ with $i=1,2,3$ and $P,Q = \pi, K, \eta$.

Pion--nucleon $\sigma$-term:
\begin{eqnarray}
\epsilon_{1,\sigma}^\pi & = & 2\epsilon_{1,N}^\pi + \frac{1}{2}
\epsilon_{2,N}^{\pi K} + \epsilon_{3,N}^\pi + \epsilon_{4,N}^{\pi\pi}
+ \frac{1}{2}\epsilon_{4,N}^{\pi K} +
\frac{1}{3}\epsilon_{4,N}^{\eta\pi} \, \, , \nonumber \\
\epsilon_{1,\sigma}^K & = & \epsilon_{1,N}^K + 
\epsilon_{2,N}^{\pi K} + \frac{1}{2}\epsilon_{3,N}^K + 
\frac{1}{2}\epsilon_{4,N}^{KK}+ \epsilon_{4,N}^{\pi K} +
\frac{1}{3}\epsilon_{4,N}^{\eta\pi} \, \, , \nonumber \\
\epsilon_{1,\sigma}^\eta & = & \frac{2}{3}\epsilon_{1,N}^\eta + \frac{1}{3}
\epsilon_{3,N}^{\eta} + \frac{1}{3}\epsilon_{4,N}^{\eta \eta} \,\,\, \,\,\, ,
\epsilon_{2,\sigma}^\pi  =  2\epsilon_{3,N}^\pi + 2 \epsilon_{4,N}^{\pi\pi} 
+ \frac{1}{2}\epsilon_{4,N}^{\pi K} \, \, , \nonumber \\
\epsilon_{2,\sigma}^K & = & \epsilon_{3,N}^K + 
\epsilon_{4,N}^{K \pi} + \epsilon_{4,N}^{K K} \,\,\,\,  ,
\epsilon_{2,\sigma}^\eta  =  \frac{2}{3}\epsilon_{3,N}^\eta 
+ \frac{2}{3}\epsilon_{4,N}^{\eta\eta}+ \epsilon_{4,N}^{\eta \pi} 
+ \frac{1}{2}\epsilon_{4,N}^{\eta K} \, \, , \nonumber \\
\epsilon_{3,\sigma}^{K \pi} & = & \epsilon_{4,N}^{\pi K} \, \, , \, \, 
\epsilon_{3,\sigma}^{\pi K} = \frac{1}{2}\epsilon_{4,N}^{K\pi} \, \, ,\, \, 
\epsilon_{3,\sigma}^{\pi \eta} = \frac{1}{3}\epsilon_{4,N}^{\eta\pi} 
\, \, , \, \, 
\epsilon_{3,\sigma}^{K \eta} = \frac{1}{3}\epsilon_{4,N}^{\eta K} 
\, \,\,  .
\end{eqnarray}
Notice that the $\epsilon_{(2,3)\sigma}$ and the
$\epsilon_{(3,4)N}$ include the factor 
$1/\Lambda^2_\chi$ which is not made explicit here.

Strangeness fraction:
\begin{eqnarray}
\epsilon_{1,y}^\pi & = & \frac{1}{2}\epsilon_{2,N}^{\pi K} 
+ \frac{1}{2}\epsilon_{4,N}^{K \pi} + \frac{2}{3}\epsilon_{4,N}^{\eta\pi} 
\, \,  , \, \,  
\epsilon_{1,y}^K = \epsilon_{1,N}^{K} 
+ \frac{1}{2}\epsilon_{3,N}^{K} + \frac{2}{3}\epsilon_{4,N}^{\eta K} 
\, \,  , \, \,  
\nonumber \\
\epsilon_{1,y}^\eta & = & \frac{4}{3}\epsilon_{1,N}^\eta +
\frac{2}{3}\epsilon_{3,N}^\eta \, \, , \, \, 
\epsilon_{2,y}^\pi = 0 \, \, , \, \, 
\epsilon_{2,y}^K = \epsilon_{3,N}^K  \\
\epsilon_{2,y}^{\eta} & = & \frac{4}{3}\epsilon_{3,N}^{\eta} 
+ \frac{1}{2}\epsilon_{4,N}^{\eta K} \, \, ,\, \, 
\epsilon_{3,y}^{\pi K} = \frac{1}{2}\epsilon_{4,N}^{K\pi} \, \, , \, \, 
\epsilon_{3,y}^{K \pi} = 0 \, \, , \, \, 
\epsilon_{3,y}^{\pi \eta} = \frac{2}{3}\epsilon_{4,N}^{\eta\pi} 
\, \,  ,  \, \, 
\epsilon_{3,y}^{K \eta} = \frac{2}{3}\epsilon_{4,N}^{\eta K} 
\, \, . \nonumber
\end{eqnarray}


\setcounter{equation}{0}
\section{Determination of the Roper--octet 
coupling constants $D_R$ and $F_R$}
\label{app:pararoper}
 
The only listed deacys to determine the coupling constants $D_R$ and
$F_R$ are the $N^* (1440) \to N \pi$ and the $\Lambda^* (1600) \to
\Sigma \pi$. Consider first the $\Lambda^*$ decay. The width follows
via
\begin{equation}
\Gamma = \frac{1}{8 \pi m_R^2} \, |\vec{q \,}_\pi| \, |{\cal T}|^2
\quad ,
\end{equation}
with $|\vec{q \,}_\pi| =334$~MeV the pion momentum in the restframe of
the $\Lambda^*$ and we use the spinor normalization $\bar{u}(p) u(p) =
2 m_R$. Straightforward calculation gives
\begin{eqnarray}
\Gamma &=& \frac{D_R^2}{16 \pi m_{\Lambda^*} F_\pi^2} \, |\vec{q
  \,}_\pi| \, \biggl[ 2E_\pi \, (E_\Sigma E_\pi + \vec{q \,}_\pi^2) - M_\pi^2 
\, (E_{\Sigma} + m_\Sigma) \biggr] \nonumber \\
& =& 149 \, \, {\rm MeV} \, \, D_R^2 \quad ,
\end{eqnarray}
with $E_\pi = (M_\pi^2 + \vec{q \,}_\pi^2)^{1/2}$ and
$E_\Sigma = (m_\Sigma^2 + \vec{q \,}_\pi^2)^{1/2}$. From the PDG value
$\Gamma (\Lambda^* \to \Sigma \pi) = (150 \pm 100)\, {\rm MeV} \cdot
(0.35 \pm 0.25) = 52.5$~MeV we
derive
\begin{equation}
D_R   =  0.60 \pm 0.41 \quad .
\label{DRval}
\end{equation}
The sign of $D_R$ is, of course, not fixed. We will chose it to be
positive in accordance with the  ground state octet $D$ coupling.
The $N^*(1440) N \pi$ effective Lagrangian is
\begin{eqnarray}
{\cal L}_{N^* N \pi} & = & \frac{g_A}{4} \, \sqrt{R} \, \bar{\Psi}_{N^*}
  \, \gamma_\mu \gamma_5 \, u^\mu \, \Psi_N + {\rm h.c.} \, \, ,
\nonumber \\g_A \, \sqrt{R} & = & D_R + F_R \quad .
\end{eqnarray}
Here, $\sqrt{R} = 0.53 \pm 0.04$ and $g_A = 1.33$ as given by the 
Goldberger--Treiman relation. As explained in some detail in 
Ref.\cite{bkmpipin}, we use the width of the $N^* $
determined from the speed plot, not the the  model--depedent
Breit--Wigner fits, $\Gamma_{\rm tot} = 160 \pm 40$ MeV. Assuming
$F_R$ to be positive (as is the equivalent hyperon coupling),
we find by using Eq.(\ref{DRval}),
\begin{equation}
 F_R = 0.11 \mp 0.41 \quad .
\end{equation}
These are the values of the Roper couplings used in the main text.

\vfill


\newpage

\section*{Tables}

\vskip 0.5in

\renewcommand{\arraystretch}{1.5}

\begin{table}[h]
  \begin{tabular}{|c|cccc|}
    \hline
    B &  $\lambda_B$ [GeV$^{2}$] 
      & $D^\beta_B$ [GeV$^{2}$] & $D^\delta_B$  [GeV$^{4}$]
      & $D^\epsilon_B$  [GeV$^{2}$]  \\  
    \hline
    $N$        & 0.030    & ($-$1.165,+0.233) & ($-$0.123,$-$0.079) 
               & (0.114,0.029) \\ 
    $\Lambda$  & 0.065    & ($-$2.013,$-$0.783) & ($+$0.077,$-$0.053)   
               & (0.219,0.035) \\   
    $\Sigma $  & 0.058    & ($-$3.626,$-$0.631) & ($-$0.802,$-$0.186)
               & (0.506,0.058) \\   
    $\Xi$      & 0.096    & ($-$3.205,$-$1.648) & ($-$0.390,$-$0.108) 
               & (0.471,0.052) \\  
    \hline
  \end{tabular}
  \medskip 
  \caption{Numerical values of the state--dependent  coefficients
           in Eq.(\protect\ref{massreso}). 
           The first and second term in the brackets 
           refers to the decuplet and Roper-octet contribution, 
           respectively.}
\end{table}


\renewcommand{\arraystretch}{1.5}

\begin{table}[h]
  \begin{tabular}{|c|ccccccc|}
    \hline
    & $b_0$ & $b_D$ & $b_F$ & $b_1$ & $b_2$ & $b_3$ & $b_8$  \\  
    \hline
    $\beta_R =0$  
                 & $-$0.606 & 0.079  &  $-$0.316
                 & $-$0.004 & $-$0.187 & $+$0.018 & $-$0.109 \\  
    \hline
    Ref.\cite{norb} & $-$0.493 & 0.066  & $-$0.213
                  & $+$0.044 & $-$0.145 & $-$0.054 & $-$0.165 \\
    \hline
    Ref.\cite{bkmz} & $-$0.750 & 0.016 & $-$0.553 & - & - & - & -  \\
    \hline
  \end{tabular}
  \medskip 
  \caption{Low--energy constants from ${\cal L}_{\phi B}^{(2)}$
            in GeV$^{-1}$ for $\beta_R=0$.\hfill }
\end{table}

\renewcommand{\arraystretch}{1.5}

\begin{table}[bht]
  \begin{tabular}{|c|ccc|ccccc|}
    \hline
      $\lambda$ & $b_0$ & $b_D$ & $b_F$ & $\krig m$  
      & $m_N^{(2)}$ &  $m_N^{(4)}$ & $\hat \sigma$  & $y$ \\
      $$[GeV] & [GeV$^{-1}$] & [GeV$^{-1}$] & [GeV$^{-1}$] 
      & [MeV] & [MeV] & [MeV] & [MeV] &  \\
    \hline
      0.8 & $-$0.643 & 0.085 & $-$0.369 & 849.& 244. & 119.& 49.4 & $-$0.10 \\
      1.0 & $-$0.606 & 0.079 & $-$0.316 & 767. & 261. & 183. & 35.6 & +0.21  \\
      1.2 & $-$0.574 & 0.072 & $-$0.281 & 613. & 266. & 232. & 25.9 & +0.42 \\
    \hline
  \end{tabular}
  \medskip 
  \caption{Theoretical uncertainties due to the  renormalization scale
    $\lambda$. For comparison, we give the second and fourth order
    contribution to the nucleon mass, $m_N^{(2)}$ and
    $m_N^{(4)}$, respectively. The third order contribution is
    $m_N^{(3)} = -272 \,$MeV. We set $\beta_R =0$.}
\end{table}

\newpage

\renewcommand{\arraystretch}{1.5}

\begin{table}[bht]
  \begin{tabular}{|c|ccc|ccccc|}
    \hline
      $\sigma_{\pi N}(0)$ & $b_0$ & $b_D$ & $b_F$ & $\krig m$  
      & $m_N^{(2)}$ &  $m_N^{(4)}$ & $\hat \sigma$  & $y$ \\
      $$[MeV] & [GeV$^{-1}$] & [GeV$^{-1}$] & [GeV$^{-1}$] 
      & [MeV] & [MeV] & [MeV] & [MeV] &  \\
    \hline
      35 & $-$0.508 & 0.082 & $-$0.319 & 910.& 163. & 139.& 37.4 & $-$0.07  \\
      45 & $-$0.606 & 0.079 & $-$0.316 & 767. & 261. & 183. & 35.6 & +0.21  \\
      55 & $-$0.691 & 0.074 & $-$0.310 & 625. & 360. & 226. & 37.1 & +0.34 \\
    \hline
  \end{tabular}
  \medskip 
  \caption{Theoretical uncertainties due to the pion--nucleon 
    $\sigma$--term. For comparison, we give the second and fourth order
    contribution to the nucleon mass, $m_N^{(2)}$ and
    $m_N^{(4)}$, respectively. The third order contribution is
    $m_N^{(3)} = -272 \,$MeV. We set $\beta_R =0$.}
\end{table}

\vskip 1.2in

\section*{Figure captions}

\begin{enumerate}

\item[Fig.1] One--loop graphs with exactly one insertion from
${\cal L}^{(2)}_{\phi B}$ (circlecross). Goldstone boson renormalizations
are not shown.

\item[Fig.2] Contribution to the scalar form factor. The
 double--line denotes the insertion of $\hat m (\bar u u + \bar d
 d)$.

\item[Fig.3] Baryon resonance excitation involving pion
loops. The double--line
represents the decuplet or the even-parity Roper octet.
Solid and dashed lines denote the ground state octet baryons and 
Goldstone boson fields, respectively.

\item[Fig.4] Scalar meson  excitation. The double--line
represents the scalars  and the circlecross a quark mass insertion
$\sim \chi_+$.

\item[Fig.5] Baryon resonance excitation: Two--loop graphs.
 The double--line represents the decuplet or the even-parity Roper
 octet or a combination thereof. Solid and dashed lines denote 
 the ground state octet baryons and  Goldstone boson fields,
 respectively. Graphs of the same topologies with one or two
 groundstate propagators are not shown.

\end{enumerate}

\newpage


\begin{center}

$\,$\vskip 1.0cm

\hskip 0.3in
\epsfysize=2.5in
\epsffile{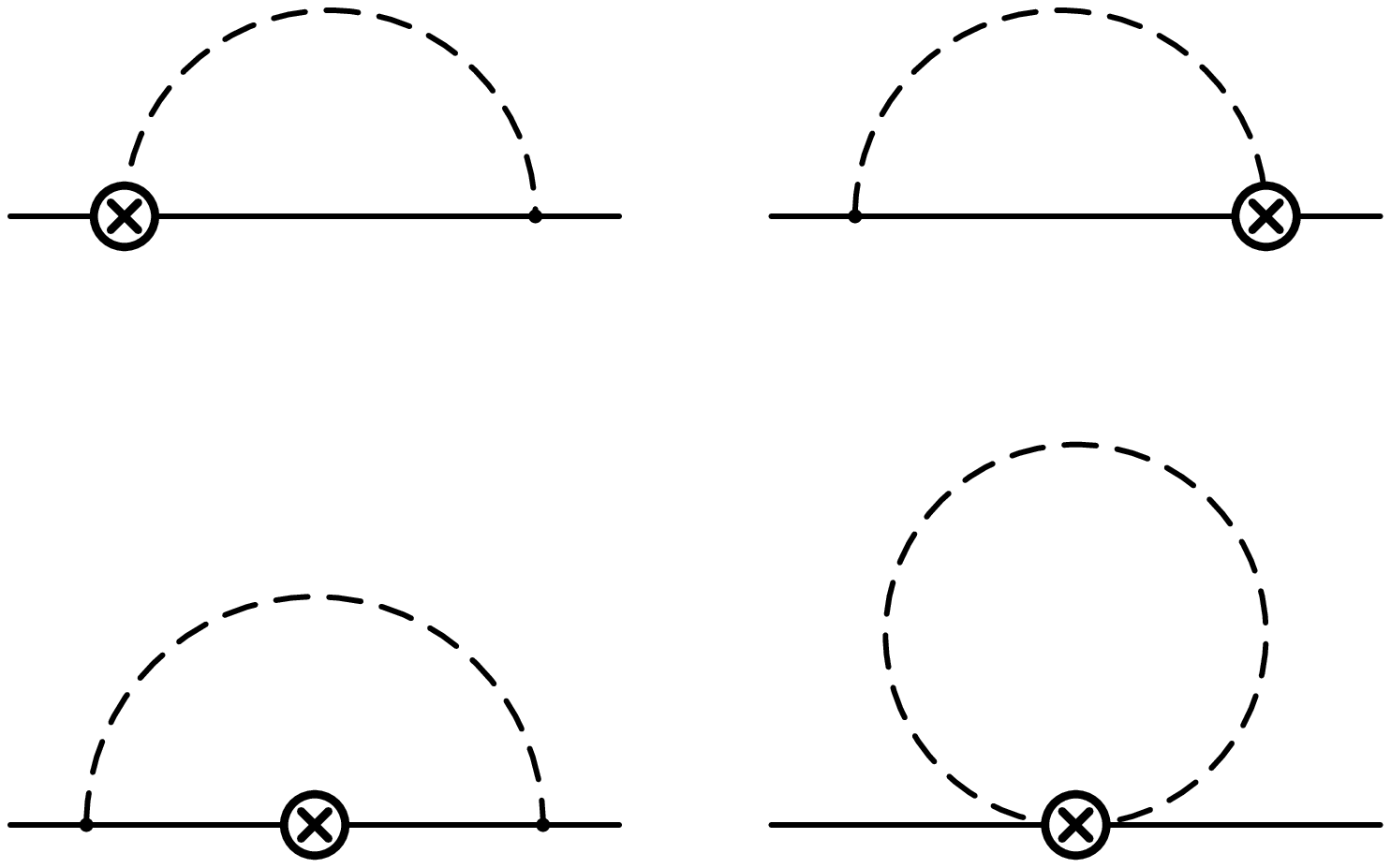}

\vskip 0.7cm

Figure 1

\vskip 1.5cm


\hskip .3in
\epsfysize=1.in
\epsffile{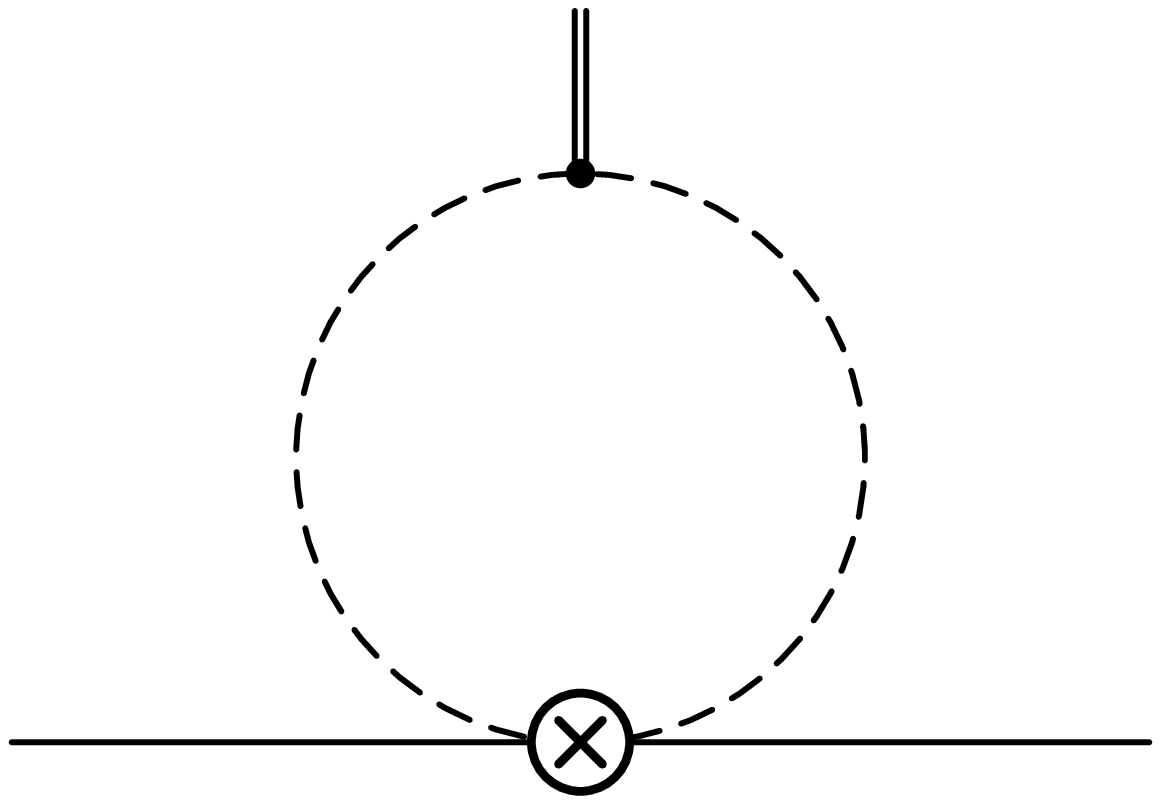}

\vskip 0.7cm

Figure 2
 
\vskip 1.5cm


\hskip .3in
\epsfysize=1.5in
\epsffile{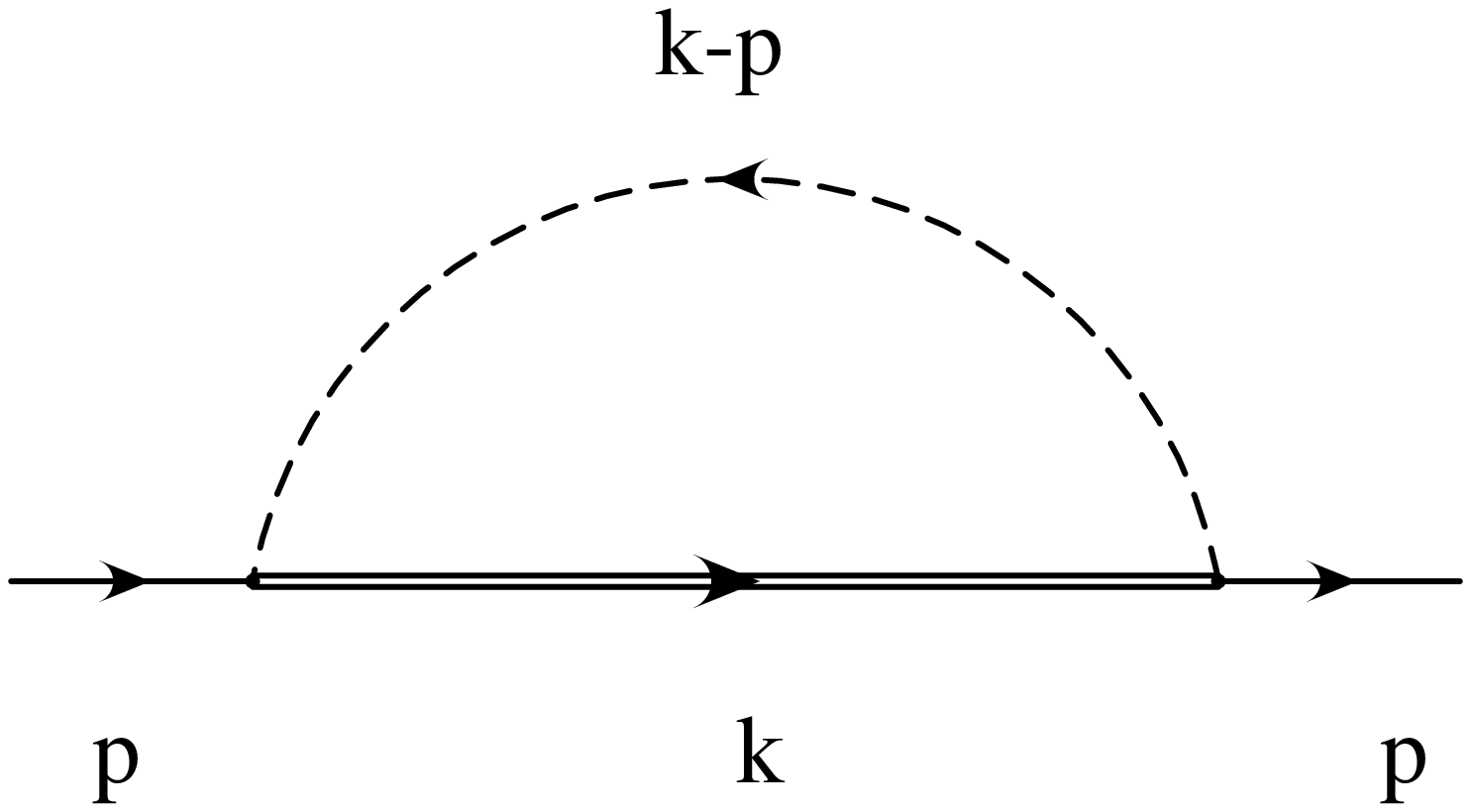}

\vskip 0.7cm

Figure 3

\vskip 1.5cm

\newpage

$\,$\vskip 1cm

\hskip .3in
\epsfysize=1.2in
\epsffile{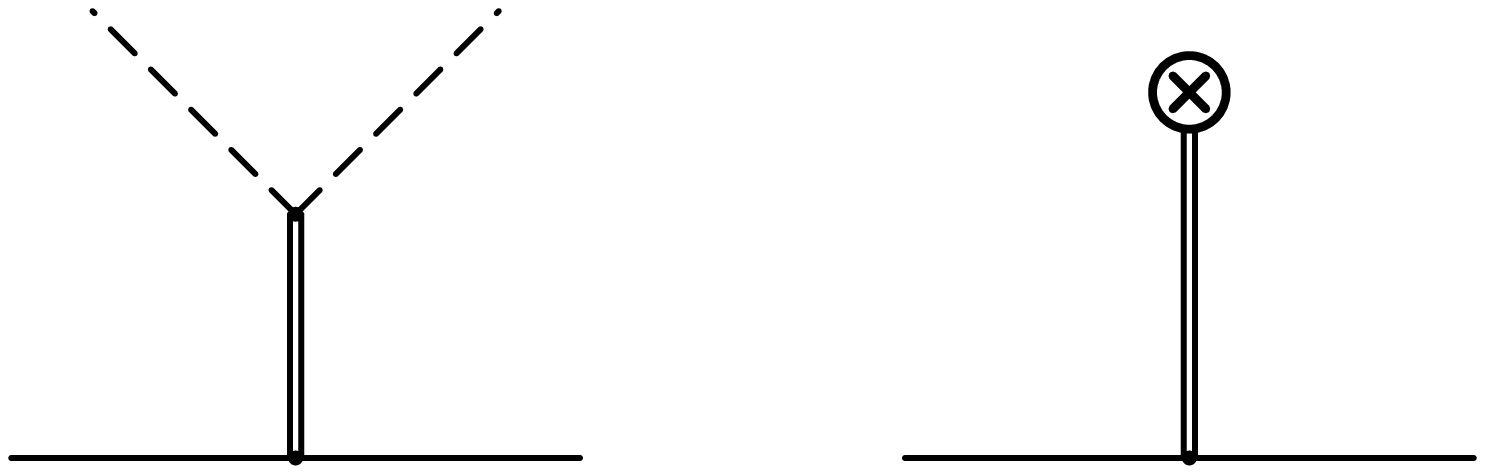}

\vskip 0.7cm

Figure 4

\vskip 2.5cm


\hskip 1in
\epsfysize=4in
\epsffile{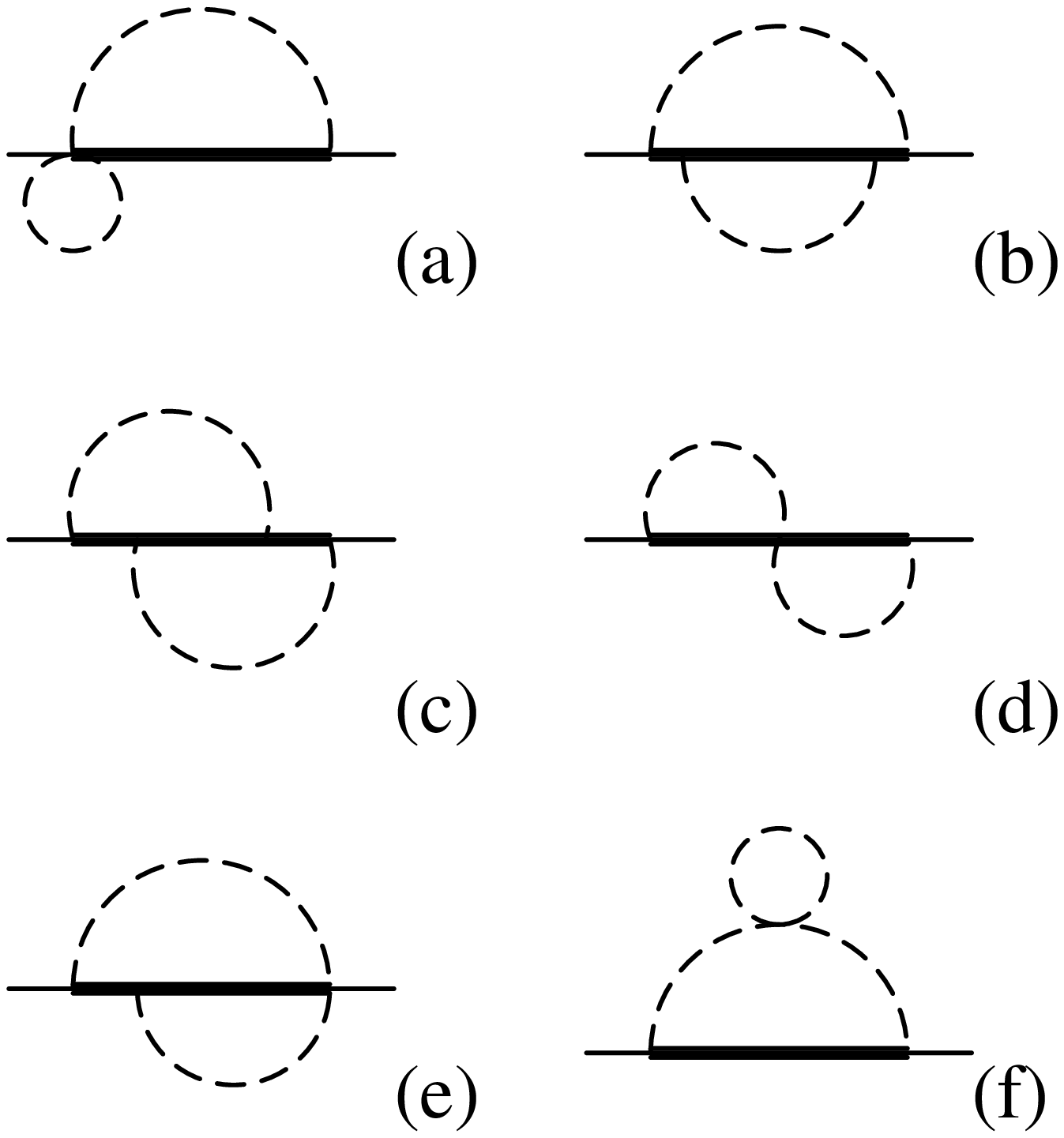}

\vskip 0.7cm

Figure 5

\vskip 1.5cm

\end{center}

\end{document}